\documentclass[11pt]{article}

\usepackage{authblk}
\usepackage{amsmath}
\usepackage{amssymb}
\usepackage{amsthm}
\usepackage{amsthm}
\usepackage{delarray}
\usepackage{amsfonts}
\usepackage{bm}
\usepackage{algorithm}
\usepackage{algorithmic}
\usepackage{eurosym}
\usepackage{graphicx}
\usepackage{multirow}
\usepackage{rotating}
\usepackage{subfigure}
\usepackage{placeins}
\usepackage{xspace}
\usepackage{pgffor}
\usepackage{url}
\usepackage{todonotes}
\usepackage[all]{nowidow}
\usepackage{fullpage}
\usepackage[authoryear]{natbib}
\usepackage{hyperref}

\newcommand{\name}[1]{\mbox{#1}\xspace}

\newcommand{\PERF}{Efficiency\xspace}
\newcommand{\Perf}{efficiency\xspace}
\newcommand{\SCAL}{Scalability\xspace}
\newcommand{\Scal}{scalability\xspace}
\newcommand{\scip}{\name{SCIP}}
\newcommand{\soplex}{\name{SoPlex}}
\newcommand{\cplex}{\name{CPLEX}}
\newcommand{\gurobi}{\name{Gurobi}}
\newcommand{\xpress}{\name{Xpress}}
\newcommand{\ug}{\name{UG}}
\newcommand{\pipssbb}{\name{PIPS-SBB}}
\newcommand{\symphony}{\name{SYMPHONY}}

\newcommand{\parascip}{\name{ParaSCIP}}
\newcommand{\alps}{\name{ALPS}}

\newcommand{\coma}{\name{Commercial 1}}
\newcommand{\comb}{\name{Commercial 2}}
\newcommand{\comc}{\name{Commercial 3}}

\newcommand{\mip}{MILP\xspace}

\newcommand{\lp}{LP\xspace}

\newcommand{\miplib}{MIPLIB\xspace}

\newcommand{\pdi}{primal-dual integral\xspace}
\newcommand{\Pdi}{Primal-dual integral\xspace}

\newcommand{\wct}{wall clock time\xspace}
\newcommand{\Wct}{Wall clock time\xspace}

\newcommand{\Z}{\mathbb{Z}}

\newcommand{\bnb}{branch-and-bound\xspace}
\newcommand{\Bnb}{Branch-and-bound\xspace}

\newcommand{\noprint}[1]{}

\newcommand{\mycitet}[2]{\citet{#2}}
\newcommand{\mycitep}[1]{\citep{#1}}

\renewcommand{\Re}{\mathbb{R}}

\graphicspath{{figures/}}

\newtheorem{definition}{Definition}

\newcommand{\PAPERTITLE}{Assessing the Effectiveness of (Parallel)
Branch-and-bound Algorithms}
\newcommand{\AUTHORSM}{Stephen J. Maher}
\newcommand{\EMAILSM}{s.j.maher@exeter.ac.uk}
\newcommand{\THANKSSM}{\thanks{\texttt{\EMAILSM}}}
\newcommand{\AFFILSM}{College of Engineering, Mathematics and Physical Sciences,
University of Exeter, Exeter, United Kingdom}
\newcommand{\AUTHORTR}{Ted K. Ralphs}
\newcommand{\EMAILTR}{ted@lehigh.edu}
\newcommand{\THANKSTR}{\thanks{\texttt{\EMAILTR}}}
\newcommand{\AFFILTR}{Dept of Industrial and Systems Engineering, Lehigh
University, USA} 
\newcommand{\AUTHORYS}{Yuji Shinano}
\newcommand{\EMAILYS}{shinano@zib.de}
\newcommand{\THANKSYS}{\thanks{\texttt{\EMAILYS}}}
\newcommand{\AFFILYS}{Dept of Mathematical Optimization, Zuse Institute Berlin,
Takustr. 7, 14195 Berlin, Germany}
\newcommand{\PAPERABSTRACT}{
  \noindent Empirical studies are fundamental in assessing the effectiveness
  of implementations of branch-and-bound algorithms. The complexity of such
  implementations makes empirical study difficult for a wide variety of
  reasons. Various attempts have been made to develop and codify a set of
  standard techniques for the assessment of optimization algorithms and their
  software implementations; however, most previous work has been focused on
  classical sequential algorithms. Since parallel computation has become
  increasingly mainstream, it is necessary to re-examine and modernize these
  practices. In this paper, we propose a framework for assessment based on the
  notion that \emph{resource consumption} is at the heart of what we generally
  refer to as the ``effectiveness'' of an implementation. The proposed
  framework carefully distinguishes between an implementation's
  baseline \emph{\Perf}, the efficacy with which it utilizes a fixed
  allocation of resources, and its \emph{\Scal}, a measure of how the
  efficiency changes as resources (typically additional computing cores) are
  added or removed. Efficiency is typically applied to sequential
  implementations, whereas scalability is applied to parallel
  implementations. \PERF and \Scal are \emph{both} important contributors in
  determining the overall effectiveness of a given parallel implementation,
  but the goal of improved efficiency is often at odds with the goal of
  improved scalability. Within the proposed framework, we review the
  challenges to effective evaluation and discuss the strengths and weaknesses
  of existing methods of assessment.
}
\newcommand{\PAPERKEYWORDS}{effectiveness, efficiency, performance, scalability, algorithm assessment,
   branch-and-bound, benchmarking}

\begin{document}


\title{\PAPERTITLE}

\author{\AUTHORSM\THANKSSM}
\affil{\AFFILSM}
\author{\AUTHORTR\THANKSTR}
\affil{\AFFILTR}
\author{\AUTHORYS\THANKSYS}
\affil{\AFFILYS}
\titlepage

\maketitle

\begin{abstract}

\PAPERABSTRACT


\end{abstract}


\section{Introduction}
\label{sect:intro}

This paper considers the challenging question of how to assess and compare the
widely differing implementations of what is perhaps the most well-known
algorithmic paradigm for solving NP-hard discrete optimization
problems---the \bnb algorithm. We introduce a general framework for assessing
the effectiveness of \bnb-based algorithms that is broadly applicable but with
a particular focus on algorithms for solving difficult discrete optimization
problems. We indicate how the current methodology for assessment fits into our
framework and also review some of the challenges in performing such assessment.
To this end, we highlight some ways that existing methods are not well-suited
for measuring aspects of effectiveness, particularly in the case of parallel
implementations.

As many terms are overloaded or inconsistently used in the literature when
discussing the assessment of implementations one of the first challenges is to
develop a clear and consistent terminology. In
Section~\ref{sect:performanceAndScalability}, we introduce a formal language for
describing the measurable properties of algorithms that we use to assess them.
Throughout the paper, we use the term \emph{effectiveness} in a general way to
denote the overall combination of these measurable properties that leads to
their observed behavior in practice. Other, more common terms related to the
assessment of implementations will be used in a more technical way.

Although \bnb is often described as an ``algorithm,'' its general
description leaves the implementation of several critical procedures
unspecified. As such,  \bnb is more a general algorithmic
\emph{framework} than a complete algorithm. In the decades since the first
variant was described by~\mycitet{Land and Doig}{LanDoi60}, a vast number of
adaptations have subsequently been presented in the literature, making it one
of the most widely implemented algorithmic approaches for solving difficult
discrete optimization problems.

The flexibility of the basic framework gives it tremendous power to solve a wide
variety of problems. Unfortunately, this power leads to a level of
sophistication of implementations that makes rigorous assessment and comparison
much more challenging.  The comparison of differing implementations has
traditionally been approached by assessing effectiveness on a fixed set of test
instances, which must be carefully chosen in order to allow a fair comparison
(or even to make comparison possible at all).  Unfortunately, in some important
cases, the use of a fixed test set for purposes of comparison is simply not
feasible using traditional measures of effectiveness. In particular, when
comparing parallel scalability, this method of assessment requires not only that
instances are solvable by all implementations on a single (or a small number) of
cores\footnote{We use this term throughout to informally refer to units of
hardware capable of doing a single sequential computation.}, but also that the
instances are not ``too easy.'' Obtaining a set of such test instances large
enough to draw valid and generalizable conclusions is difficult, especially when
comparing implementations with contrasting strengths.

Generally speaking, there are two substantially different approaches to the
assessment of algorithms: theoretical and empirical. Theoretical analysis has
many advantages, most notably the rigor of the underlying analysis. It also
eliminates many of the difficulties and confounding factors related to
properties of the hardware, programming language used, compiler, programmer
skill, and other extraneous environmental factors that interfere with analysis
of the algorithm itself.

Unfortunately, in the case of discrete optimization, the tools for theoretical
analysis that currently exist are largely inadequate for the task. The
worst-case measure by which implementations are traditionally judged in a formal
complexity analysis (due to the difficulties associated with doing an
average-case or other types of analysis) is not capable of distinguishing
between variants of \bnb, all of which are theorized at present to have the same
or similar worst-case behavior. That leaves little option but to compare
implementations empirically.

Within the empirical realm, computational experimentation is the main tool
employed by researchers for assessing the relative effectiveness of both
algorithms and software. While there has long been a heavy reliance on
computational experiments, relatively little effort has gone into the
development of a rigorous science of experimental analysis for algorithms.
While some best practices have been discussed, much of the early work is now
dated and newer technology calls for updated techniques that unify concepts of
analysis for both sequential and parallel implementations. In what follows, we
propose a unified framework for assessing the effectiveness of algorithms.

\subsection{Basic Setting}

A (mathematical) optimization problem is that of determining
\begin{equation} \label{eq:opt}
z^* = \min_{x \in \mathcal{F}} f(x),
\end{equation}
where $f: \Re^n \rightarrow \Re$ is a given objective function and
\begin{equation*}
\mathcal{F} = \{x \in \Z^p \times \Re^{n-p} \mid g_i(x) \leq
b_i, 1 \leq i \leq m\}
\end{equation*}
is the feasible set. We assume for convenience that $f$ is continuous and
$\mathcal{F}$ is compact so that the minimum exists. A \emph{class} of such
optimization problems is a parametric family in which the feasible regions and
objective functions have a certain well-defined structure, i.e., the
functions, $f(x)$ and $g_{i}(x)$, are from a particular class (e.g., linear
functions) that are parameterized on certain input data. In such a case,
specific values of the input parameters correspond to a
specific \emph{instance} in the class. Typically, a class contains instances
with arbitrarily large numbers of variables and constraints.

Discrete optimization problems can be roughly defined as those for which at
least some fixed proportion of the variables must take integer values in all
feasible solutions. Such problems are particularly amenable to solution by a
search algorithm, such as \bnb. We do not describe branch-and-bound itself in
detail, since it is well-described in many other places. For a formal
description of \bnb in the context of solving discrete optimization problems,
including an introduction to the issues arising in its parallelization, we
refer the reader to~\mycitet{Ralphs et al.}{RalShiBerKoc18}.

We consider only \emph{exact} algorithms. Such algorithms produce both an
optimal solution $x^* \in \operatorname{argmin}_{x \in \mathcal{F}} f(x)$
to~\eqref{eq:opt} (when required) \emph{and} a proof of the optimality of that
solution (or a proof that some alternative termination criteria have been
satisfied). This proof usually comes in the form of a \emph{\bnb tree}
(see~\cite{GuzRal07} for details on how such a tree encodes a proof) that
provides global upper and lower bounds on $z^*$ (upper and lower bounds are more
generally referred to as \emph{primal} and \emph{dual}, respectively, to allow
for the possibility that the sense of the optimization is maximization).  These
bounds are updated throughout the solution process and the algorithm usually
terminates when they are (approximately) equal. In cases of early termination,
such as when reaching a time limit, the reported bounds are likely not equal but
are nevertheless valid.

A fully rigorous empirical analysis requires distinguishing between an algorithm
and its software implementation. This in turn requires distinguishing between
factors arising from i) properties of the algorithm itself (as a mathematical
abstraction); ii) incidental properties of the software implementation
(compiler, language, operating system kernel, etc.); and iii) incidental
properties of the hardware on which the experimentation is done (memory
hierarchy, network topology, etc.). Such an analysis would obviously be
painstaking and impractical, but it's still important to acknowledge that the
inability to account for these confounding variables is a significant
limitation.  An accounting of the full range of confounding variables is beyond
the scope of this paper.
In acknowledgment of these challenges, we have, however, endeavored to use
the term \emph{implementation} whenever we are speaking of the assessment of a
particular algorithm, since it is the implementation, not the algorithm
itself, that is actually being assessed.

There are innumerable practical aspects that we do not fully address in this
work. It is unavoidable that different implementations may produce different
solutions and different proofs for the same problem---this is one of the
difficult challenges we face. Most implementations do not actually output the
proof they construct (parts of the proof that are no longer relevant are
typically discarded to save memory), but rather only output the upper and
lower bounds. Finally, as the computations are done in finite precision
arithmetic, the proofs produced are only approximate (see,
e.g.,~\mycitep{Lil04}). While numerical issues arising from the use of
floating point computation and other practical limitations of typical
implementations are important in the study of these algorithms, they are less
important in practice and have little impact on the development of the
concepts presented here. We therefore avoid unnecessary discussion of these
details.

\subsection{Previous Work \label{sect:literatureReview}}

Assessing the relative effectiveness of algorithms for solving mathematical
optimization problems and their software implementations dates back as far as
the earliest packages for solving such problems. Since the early days of
computing, scores of papers have endeavored to present comparisons of the
software implementations of various algorithms. One of the first efforts to
systematically evaluate optimization software is presented by~\mycitet{Hoffman
et al.}{hoffman53} in the context of evaluating solvers for linear
optimization problems (LPs). Due to the inherent limitations of theoretical
analysis of these algorithms, such empirical comparisons have been crucial in
assessing progress in the field and in evaluating the merit of different
algorithmic approaches. In the decades since, computational research has
become mainstream and an enormous number of papers have been published that
touch on the issues surrounding empirical research in general. We do not
attempt to survey this entire literature, but rather refer the reader to the
general overview given by~\mycitet{Lilja}{Lil04}. In what follows here, we
briefly review the most closely related works from the optimization
literature.

The dependence on empirical assessment for research in optimization algorithms
has prompted many calls for improved standards and guidelines. One of the
early works suggesting standards for computational experiments was
from~\mycitet{Ignizio}{ignizio71}, who argues that such standards are required
to maintain the high status of research in the field. The need for standards
has lead to various attempts at laying out guidelines to direct researchers in
performing computational experiments. One set of recommendations are provided
by~\mycitet{Crowder et al.}{crowder78}, covering the experimental design, the
selection of test instances, and the possible measures for comparing
algorithms. Building upon the work of~\mycitet{Crowder et
al.}{crowder78}, \mycitet{Jackson et al.}{jackson90} presents an updated set
of guidelines for the reporting of computational results. While the guidelines
presented by~\mycitet{Jackson et al.}{jackson90} are not prescriptive, they do
highlight a number of factors that should be considered when conducting
computational experiments. Following the report on the guidelines
by \mycitet{Jackson et al.}{jackson90}, a number of suggestions about conducting
experiments have also been made by \mycitet{Greenberg}{Gre90}.  Later, another
survey of many recommended best practices for experimental analysis was given
by~\mycitet{Johnson}{Joh02}.

The heavy reliance on empirical assessment in computational mathematical
optimization has lead to a number of proposals for increasing the rigor with
which such analysis is done, but these have not met wide adoption. One of the
most comprehensive proposals for a more scientific approach to the empirical
analysis of algorithms was developed by~\mycitet{McGeoch}{McG87} in her
dissertation (and in a later book~\mycitep{McG12}). This was followed by a
call by~\mycitet{Hooker}{hooker94} for an empirical science of algorithms and
a call by~\mycitet{Moret}{MorSha01,Mor02} for the development of a discipline
of experimental algorithmics. This more scientific approach is to be set apart
from the more theoretical aspects of the analysis of algorithms, with a
particular focus on the evaluation of practical implementations.
The difficulties associated with computational experiments are further
discussed in a subsequent paper by~\mycitet{Hooker}{hooker95} with a focus on
the evaluation of heuristic methods. Further,~\mycitet{Coffin and
Saltzman}{CofSal00} proposed methods for statistically analyzing experimental
data. 

More recently, a review of the best practice for comparing optimization
algorithms is presented by~\mycitet{Beiranvand et al.}{beiranvand17}; however,
the discussions are primarily focused on continuous, unconstrained
optimization algorithms. A more in-depth discussion related to guidelines for
computational experiments is presented by~\mycitet{Kendall et al.}{kendall16}.
The focus of~\mycitet{Kendall et al.}{kendall16} is the development of good
laboratory practice for performing optimization-based research. An important
topic relevant to discrete optimization algorithms that is discussed
by~\mycitet{Beiranvand et al.}{beiranvand17} is the presentation of empirical
data---particularly methods for visualizing the results of computational
experiments. There has been some limited work done on this topic over time
(see~\mycitep{CofSal00,Joh02,San02,dolan02}), but more focus is needed on this
often overlooked yet important topic.

Another topic that deserves more focused attention is the development of
suitable objective measures for assessing algorithm effectiveness.
While~\mycitet{Crowder et al.}{crowder79} covers all aspects of algorithm
evaluation, the literature on current practices regarding measures of
effectiveness is limited. Primarily,~\mycitet{Crowder et al.}{crowder79}
examines the classical measures of CPU time and iteration counts.
Additionally,~\mycitet{Crowder et al.}{crowder79} propose the \emph{observed
rate of convergence}, which is aimed at assessing the empirical effectiveness
of iterative algorithms (in contrast to the theoretical rate of convergence,
which is a worst-case measure). The discussion presented by~\mycitet{Ahuja and
Orlin}{ahuja96} reviews measures that can aid in gaining further insight into
the behavior of an algorithm. The suggestion by~\mycitet{Ahuja and
Orlin}{ahuja96} is to evaluate algorithms using representative operation
counts. This measure is aimed at controlling effects that are attributable to
irrelevant details of the implementation and have little to do with assessment
of the mathematical algorithm itself. \mycitet{Moret}{Mor02} discusses
choosing an appropriate measure, but makes no specific
recommendations. \mycitet{McGeoch}{McG87,McG12} also discussed the choice of
suitable measures in her work.

Most of the work mentioned so far has focused purely on optimization software
designed to run on a single core. The growth in multi-core computing
technology introduced a wide array of additional challenges in assessing the
effectiveness of software implementation. As a result, many papers were
produced describing not only the implementation of parallel algorithms, but
also how to measure their performance. Many of the early papers focused on the
most straightforward measures of \emph{strong \Scal} (defined in
Section~\ref{sect:scalability}), such as parallel speed-up. \mycitet{Barr and
Hickman}{barr93,barr93advance} summarize the general issues surrounding
assessment of parallel implementations and discuss how to compute parallel
speed-up, also observing that anomalous behavior, such as super-linear speed-up,
is possible in some cases. They observe that the proper evaluation of parallel
\Scal (defined later in Section~\ref{sect:scalability}) can lead to advances in
the underlying sequential implementations.

Despite the continued use of speed-up on a fixed test set as a measure
of \Scal, it was already observed in the early days of parallel computing that
this method of evaluation has many drawbacks. \mycitet{Amdahl}{amdahl67}
famously noted that as long as there is a certain fraction of the computation
that is inherently sequential, the parallel speed-up that can be achieved on a
fixed computational task will be limited, no matter how many additional
resources are utilized. This observation is easily borne out in practice. To
overcome this limitation, it is necessary to scale up the computational task
itself as the resources are scaled in order to maintain an efficient usage.
This more nuanced view was first promoted by~\mycitet{Gustafson}{gustafson88}
and was later taken up by many other authors, resulting in the development of
the notion of \emph{weak scalability}. \mycitet{Kumar and Gupta}{kuma:scala}
summarize many of the works from a large body of literature related to
assessing scalability and describe notions similar to weak scalability, though
they do not use the term. A thorough yet succinct introduction to
parallelization and different notions of scalability can also be found in
Chapter 5 of~\mycitet{Hager and Wellein}{HagWel11}. We discuss how these notions
fit into our own framework in Section~\ref{sect:scalability}.

\subsection{Contributions \label{sect:paperStructure}}

This paper presents a unified framework for assessment that incorporates
independent yet interlinked concepts of effectiveness---\emph{\Perf}
and \emph{\Scal}. These concepts link the assessment of sequential and
parallel algorithms, traditionally analyzed in different ways. At the core of
the proposed framework is an abstract concept of resource consumption based on
the following ideas.
\begin{itemize}
\item The notion that \emph{resource consumption} is at the heart of
    what is meant by ``effectiveness'' and that assessment is primarily
    concerned with measuring how efficiently resources are being used, as well
    as the trade-offs between different resources.
\item The clear separation of baseline \emph{\Perf}, applicable in an
    environment with fixed resources, and \emph{\Scal}, applicable to
    environments with dynamically allocated resources.
\item The differentiation of traditional \emph{measures of \Perf},
    from \emph{measures of progress} and \emph{measures of work}, which can be
    used as proxies and allow the comparison of computations that
    cannot be completed in a reasonable amount of time.
\item A proposal for measuring \emph{weak \Scal} (defined later) based on a
    chosen measure of progress that is more robust than the more familiar
    \emph{strong \Scal} that is typically used to assess parallel
    implementations.
\end{itemize}
Through the lens of this framework, we discuss the strengths and limitations
of existing practices for the assessment of both sequential and parallel
implementations of \bnb algorithms. 
To the best of our knowledge no recent discussion of this topic has attempted
to view assessment in this broad fashion. We hope this work contributes to the
continued development of best practice for assessment, but emphasize that it
is intended more as a starting point than a final conclusion.

A key part of this work is to highlight a fundamental and vexing challenge in
designing effective parallel algorithms: the fact that the goal of
achieving \Scal and the goal of achieving \Perf may be at odds, especially
for \bnb implementations. Better baseline \Perf very often leads to
decreased \Scal. Although baseline \Perf is generally thought of as critically
important, good \Scal matters more in the limit. Whether \Perf or \Scal is
more important in actual practice depends crucially on the environment in
which the implementation will be deployed. This is why assessing and comparing
implementations requires the separate analysis of \Perf and \Scal. It's also why
one cannot always say whether one implementation is ``better'' than another.
The measure of goodness has multiple objectives.

To aid the exposition of the concepts in this paper, examples from various
computational experiments are presented. These computational experiments
involve using various \bnb implementations---sequential and parallel---to
solve the instances from the \miplib{}2010 benchmark test
set~\mycitep{koch11}. Due to limitations of the comparison methods being
described, it is not possible to use all instances of the test set in every
example. In each of the following examples, the method for selecting the
instances is clearly described. All computational experiments have been
conducted using a cluster of Intel Xeon E5-2670 CPUs with 2.50GHz and 6.6 GB
memory per core.

\section{Assessing Effectiveness \label{sect:performanceAndScalability}}

One way of informally formulating the goal of assessment is as a methodology
for approximating the (cumulative) probability distribution function of some
measure of effectiveness across all instances in a certain class. When
comparing implementations, it is the true probability distributions that we
would ideally like to use for comparison, but because the true distributions
are unknown, we must approximate them by determining the empirical cumulative
distribution function (CDF) with respect to a smaller set of test instances
that are chosen so as to be representative of the full class (see
Section~\ref{sect:instanceSelection}). This is a rough approximation at best
and a uniform probability distribution is typically assumed for simplicity.
Solution of these instances constitute the \emph{benchmark computation}. In
general, there are two distinct steps in comparing the effectiveness of
algorithm implementations:
\begin{enumerate}
\item Measure the effectiveness of each implementation with respect to one or
more chosen measures on a set of selected instances \emph{individually}.
\item Aggregate this data for each implementation
and then produce either summary statistics or a visualization (or both) that
enables valid conclusions to be drawn about effectiveness across an entire
test set (and ultimately across an entire class of instances).
\end{enumerate}
The goal of the latter step is typically to identify which algorithm is
``better.'' This choice is obviously subjective. So the goal in summarizing
and/or visualizing is usually to enable the reader of the study to draw their
own conclusions based on the data. Any general conclusions regarding dominance
of one implementation over another should ideally be done using statistical
methods, as mentioned earlier, though this is difficult to do in a rigorous way.

A central aspect of our framework for assessing the effectiveness of an
implementation is the measurement of the resources required to perform the
benchmark computation. 
\begin{definition} \label{def:resource}
A \emph{resource} is an auxiliary input, some measurable quantity of which is
required to produce the result of a computation.
\end{definition}
\noindent The notion of a resource is most typically associated with hardware,
primarily compute cores and memory, but the notion is very general and can
also include more abstract things like \emph{energy}. The underlying assumption
is that the required resources have a cost and that the goal of the
computation is to minimize the overall cost. One important resource that is
not concretely associated with physical hardware is \emph{time}, the
measurement of which is both crucial and difficult. Two differing notions of
time are \emph{wall clock} and \emph{CPU} time.
\begin{definition} \label{def:wallclock-time}
\emph{Wall clock time} is a measure of the amount of real time
required to perform a computation.  
\end{definition}
\begin{definition} \label{def:cpu-time}
\emph{CPU time} is a measure of the actual number of atomic operations
required to perform a computation, typically scaled by the amount of real time
required to perform such an operation.  
\end{definition}
\noindent In the case of a sequential computation on a single core, CPU and wall
clock time typically do not differ much and the distinction may be unimportant.
The distinction is crucial in the case of measuring the effectiveness
of parallel implementations. In parallel computations, \emph{core hours} are a
related resource that may be limited both in total quantity and in the rate at
which they are delivered. The way in which core hours are allocated to a
computation plays a rather unique role in the analysis, as further described in
Section~\ref{sect:scalability}. For now, we simply note that, assuming no other
computational tasks are competing for these resources during testing, the number
of \emph{core hours} allocated to a computation is simply the \wct multiplied by
the number of cores.  For a more detailed discussion of the details of time
measurement we refer the reader to \mycitet{Lilja}{Lil04}.

In what follows, we distinguish carefully between the two aspects of
effectiveness that were mentioned earlier. \emph{\PERF} measures the
effectiveness with which an implementation utilizes a fixed set of allocated
resources. \emph{\SCAL} considers the trade-off between resources that can be
allocated dynamically (e.g., using more cores to reduce the \wct).

The \Scal and \Perf of an implementation combine to determine how effective it
will be in practice. However, one must carefully consider the trade-off between
these when designing algorithms in modern computing environments. It is
possible for an implementation to have good baseline \Perf and poor \Scal, as
well as the reverse. An example of such behavior is
shown in Table~\ref{tab:alpsKnapsack}, which presents the wall clock execution
times for the parallel implementation of both a naive \bnb algorithm for
solving the knapsack problem, implemented using the \alps framework
(see~\mycitet{Xu et al.}{xu09} for a complete description), and the \bnb
algorithm of a representative commercial solver. 
\begin{table}[h]
  \centering
  \small
  \begin{tabular}{l|rrrrr}
    Cores & 1 & 4 & 8 & 16 & 32 \\
    \hline \alps (sec) & 132.835 & 75.133 & 43.736 & 22.212 & 13.339 \\
    \quad \emph{Speed-up} &  & 1.768 & 3.037 & 5.98 & 9.958 \\
    Commercial (sec) & 0.01 & 0.01 & 0.02 & 0.03 & 0.22 \\
    \quad \emph{Speed-up} &  & 1 & 0.5 & 0.333 & 0.045 \\
  \end{tabular}
  \caption{The solution times and parallel speed-up (see
  Section~\ref{sect:changesEfficiency}) for \alps and a commercial
  implementation when solving a knapsack problem instance.}
  \label{tab:alpsKnapsack}
\end{table}
The naive implementation, due to its lack of sophisticated algorithmic features,
requires significantly more time than the commercial implementation on a single
core but exhibits good reductions in the required time as the number of cores is
increased (i.e., good \Scal). For the commercial implementation, the addition of
more cores results in a \emph{longer} solution time, due to its excellent
baseline efficiency---making the efficient use of additional resources
difficult.  This is obviously an extreme example, but it helps illustrate the
trade-off between \Perf and \Scal.

The relative importance of \Perf and \Scal in practice is a function of a number
of factors, including, but not limited to, the environment in which the
algorithm will be deployed and the properties of the instances to be solved.
The fact that different implementations are tuned for different environments is
one of the things that makes comparison difficult. When comparing
implementations, it is important to take into account and measure these
differing properties, in order to gain more insight into the trade-offs between
them.

\subsection{\PERF \label{sect:performance}}

Building on the abstract notion of resource consumption introduced above, a
\emph{measure of \Perf} is defined as follows. 
\begin{definition} \label{def:efficiency}
A \emph{measure of \Perf} for a given benchmark computation is the amount of
one chosen resource that is required to perform that computation, with the
level of all other resources fixed.
\end{definition}
\noindent In a classical analysis, the number of cores is typically fixed and
the required \wct is the resource that is measured. This leads to the
traditional view of ``\Perf'' as a property of sequential implementations. By
allowing resources other than \wct to vary, however, we obtain alternative
measures of \Perf that can be applied to parallel algorithms. For example, we
could fix the amount of \wct and memory, asking instead how many cores a
parallel implementation would require to complete a given computation.

In cases where a particular benchmark computation cannot be fully completed with
the allocated resources, an alternative is to use a \emph{measure of progress}
or a \emph{measure of work} as a proxy.
\begin{definition} A \emph{measure of progress} is an estimate of (or proxy for)
the fraction of a benchmark computation completed by an implementation, given
a fixed bundle of resources.
\end{definition}
\noindent Measuring progress is far from straightforward in many algorithmic
contexts, especially with respect to \bnb. If a reliable measure of progress
were to exist, however, this would allow comparisons to be carried out on the
basis of efficiency without requiring a benchmark computation that could be
completed under all experimental conditions with all implementations,
addressing one of the key difficulties discussed earlier.

Unfortunately, no measure of progress proposed for \bnb to date can be
considered effective in practice. There have been a number of attempts to
develop methods of estimating tree size~\mycitep{belov17,anderson18} and these
seem to be the most obvious candidates for a measure of progress. Statistical
estimates have also been shown to be viable in some
contexts~\mycitep{anstreicher.et.al:02}. We discuss later in this paper how
the primal integral of \mycitet{Berthold}{berthold13} may hold promise in some
limited contexts.

As an alternative to measuring progress, and as a proxy for \Perf, it is
possible to directly measure the amount of work that has been performed during a
computation. Measuring the amount of work can be much easier than
measuring \Perf and progress.
\begin{definition}
  A \emph{measure of work} is a direct measure of the amount of work done in
  performing a benchmark computation, generally expressed in terms of the number
  of atomic units of computational effort executed in order to perform the
  computation.
\end{definition}
\noindent The work performed during a computation is usually defined as a count
of the number of times some set of defined atomic operations are performed. For
example, this could be the number of arithmetic operations performed, the number
of times a certain bounding subproblem is solved, or the number of \bnb nodes
enumerated.

Unfortunately, simply measuring the amount of work does not indicate how the
work contributed to the progress of the computation, i.e., whether the
work was \emph{useful}. This is particularly true in the context of \bnb
algorithms for the solution of discrete optimization problem, where the goal
is typically to limit the amount of work done that was useless in hindsight,
e.g., time spent pursuing sub-trees that turn out to contain an optimal
solution. For this reason, measures of work are not often good indicators of
efficiency and don't indicate much about progress in this context.
Nevertheless, they are useful in some contexts, especially in measuring
parallel scalability, where simply maximizing the throughput of computational
tasks can be an important goal.

\subsection{\SCAL \label{sect:scalability}}

\emph{\SCAL} is a property of an implementation
deployed in a dynamic computing environment and specifically addresses the
notion that the bundle of resources required to perform a certain benchmark
computation is not unique---there is a natural trade-off of some resources for
others. The most often analyzed trade-off is between \wct and cores.  By
increasing the number of cores by a factor of $N$, we hope that the \wct time
required to complete a given benchmark computation, such as solve an \mip
instance to optimality, can ideally be reduced by a factor of $1/N$.  In other
words, the computation should ideally be completed in the \emph{same number of
total core hours} (see Section~\ref{sect:performanceAndScalability}), no matter
how many \emph{cores} are supplied. Unfortunately, this rarely occurs in
practice. Typically, the use of more cores introduces overhead. As a result, an
increase in the number of cores also increases the total number of core hours
required to complete a computation---the computation becomes less
\emph{efficient}. 

We think of \Scal analysis abstractly in our framework as the analysis of how
some measure of \Perf changes as the bundle of resources is varied in some
way. A canonical analysis, as described above, investigates the trade-off
between cores and wall clock computing time, but other trade-offs could also
be analyzed. In the literature, the term
\emph{\Scal} may be used to refer to many different aspects of an implementation
or a hardware platform. We use it here to refer to two distinct concepts, known
as \emph{strong} and \emph{weak} \Scal, which we now formally define.

\paragraph{Strong \SCAL.} \emph{Strong \Scal} is the ability of a
particular implementation to exploit the trade-off between two resources with
respect to a single benchmark computation.
\begin{definition} \emph{Strong \Scal} is the change in the level of one
resource required to perform a benchmark computation, as a function of the
level of a second resource, with all other resources
fixed. \label{def:strongScalability}
\end{definition}
\noindent A classical (strong) \emph{\Scal analysis} involves a systematic
assessment of the trade-off between time and cores. When the number of cores
allocated to a computation that originally took \wct $T$ is increased
by a factor of $\alpha$, we ideally hope that the \wct required is reduced to
$T/\alpha$

Strong \Scal appears to be a very natural way of measuring the effectiveness
of a parallel algorithm as resources are scaled, but it has some significant
issues in practice, as mentioned earlier in
Section~\ref{sect:literatureReview}. To see why, note that all computations
involve a certain fraction that is inherently sequential (for example,
reading in the data). For a fixed test set, as the number of cores is
increased, this part of the computation eventually becomes such a high
fraction of the total that no additional scaling is possible beyond a certain
point (this is known as Amdahl's Law~\mycitep{amdahl67}). This is one of the
many challenges faced when using a single test set in performing a \Scal
analysis. Unfortunately, as we describe below, there are even more significant
challenges to be faced when \emph{not} using a fixed test set.

\paragraph{Weak Scalability.} The contrasting concept of \emph{weak \Scal} is
both more difficult to define and more difficult to measure, though it has
significant \emph{theoretical} advantages over strong \Scal. Weak \Scal measures
the effect of increasing exactly one of the available resources while
simultaneously scaling the computation itself. In other words, we do not
necessarily limit the testing to a single fixed benchmark computation. This
concept results in the slightly vague definition below, which differs slightly
from the similarly vague alternatives appearing in the literature but still
captures the essence of what is meant conceptually by weak scalability.
\begin{definition}
  \emph{Weak \Scal} is the amount of additional computation that can be done as
  the level of a single resource is increased, holding all other resources
  fixed. 
  \label{def:weakScalability}
\end{definition}
\noindent The resource being increased is usually cores. Meanwhile, the wall
clock time allocated to the computation remains constant. Hence, we are
typically measuring how much more useful computation (not just work) we can
get done with more cores, but the same amount of \wct. Although it is not
exactly clear what is meant by ``the amount of additional computation'' in the
above definition, it should be clear why measuring this kind of scalability
requires varying/scaling the computation to be done at the same time as the
resources are varied. The difficulty in rigorously defining how the
computation should be scaled and exactly what is meant by the above definition
is part of the reason why weak \Scal is a less frequently used measure.

In principle, this scaling should be done by increasing the ``difficulty'' of
the test set, e.g., to be able to say that computation A is ``twice as
difficult as'' computation B. This apparently requires producing multiple sets
of instances with different, objectively measurable levels of difficulty, such
that one can experimentally determine the most difficult set that can be
handled under given conditions.
For this reason, weak \Scal can be most easily measured for computations for
which the total amount of work required \emph{is} objectively a predictable
function of the input size. Matrix multiplication is an example application
where weak scalability can be easily assessed. The naive implementation for the
multiplication of two square matrices, each with $n^2$ elements, requires
$O(n^3)$ operations.  With two cores, if the algorithm is perfectly scalable we
can expect to do twice as many operations in the same amount of time. So we
should ideally be able to multiply two square matrices, each with $(2^{1/3}n)^2$
elements, since this would require $(2^{1/3}n)^{3} = 2n^3$ operations. If the
operations could be split evenly between the two cores without introducing any
additional work (not the typical case), then the multiplication of these larger
matrices could be accomplished in the same time on two cores as the
multiplication of the smaller matrices on a single core.  This perfect scaling
is rarely achieved and the actual size of the matrices that could be multiplied
with two cores will be somewhat below the perfect scaling factor of
$(2^{1/3}n)^2$. In this context, the weak \Scal of the parallel implementation
is measured by the difference between perfect and realized scaling.

Unfortunately, it is impossible, for all practical purposes, to objectively
measure the ``difficulty'' of a given test set a priori when dealing with an
NP-hard problem (this is the very nature of such problems). Size is well-known
not to be correlated with difficulty in solving mathematical optimization
problems; however, it is hard to imagine any other reliable measure. The time
required to perform a given computation with a baseline implementation is a
measure that may be useful in some situations, but it certainly cannot be
considered an objective measure when comparing implementations with much
different capabilities. One could imagine theoretical measures, such as the
provable minimum size of a \bnb tree obtainable under certain algorithmic
assumptions. However, such measures are generally impractical.

\paragraph{Alternative Notions.} Alternatively, we propose here some new
notions of scalability that use measures of progress in order to retain the
advantages of working with a fixed test set while addressing the need for a
dynamically scaled computation. If we could effectively measure the fraction of
a large computation that could be completed with a fixed resource configuration,
then we could scale the computation by employing larger fractions of the same
computation. We conjecture that such notions of scaling may be easier to
implement in this setting than the more conventional ones.
In particular, if we can effectively measure progress, we can consider the
trade-offs underlying the assessment of scalability from additional points of
view. For example, we might ask questions like the following.
\begin{itemize}

\item How does the fraction of a given computation we can complete in a fixed
amount of time change as we increase the number of cores?

\item How does the fraction of a given computation we can complete change as we
increase the amount of time allocated to the computation while keeping the total
    number of core hours fixed?

\end{itemize}
As we have already emphasized, applying these alternative notions of \Scal
requires a reliable measure of progress.
On the face of it, devising a reliable measure of progress would seem as
difficult as the problem of predicting how long a given computation will take,
which is also well-known to be intractable. The difference, however, is that
having performed a computation to completion once, we get information that may
enable us to measure progress with respect to the \emph{same computation}, but
executed in a different fashion experimentally (with more cores, for example).

\section{Overview of Challenges \label{sect:theoreticalEmpirical}} 

In this section, we survey some of the many challenges associated with the use
of empirical analysis as a tool for comparing implementations, in a general
sense, and \bnb in particular.

\subsection{Sophistication of Implementations \label{sect:assessmentChallenges}}

As described earlier, \bnb is not so much an algorithm as an algorithmic
framework. Modern implementations are increasingly sophisticated, which has lead
to dramatic gains in the ability to solve difficult instances. This
sophistication has also introduced much difficulty in the assessment of these
implementations---making their behavior a challenge to understand and analyze.

\paragraph{Sequential Implementations.} Sophisticated implementations of the
\bnb algorithm actually consist of collections of many different
semi-indep\-endent algorithmic components, each controlling a different aspect
of the overall algorithm. Examples of these algorithmic components include
branching methods, which determine the best branching candidate at each node of
the \bnb tree; separation routines, which generate valid inequalities that
eliminate fractional LP solutions from the feasible region; as well as primal
heuristics, presolving, and propagation routines. Among other things, these
algorithmic components are responsible for solving the many decision problems
that arise during execution.  Readers unfamiliar with the function of individual
components in a modern \bnb algorithm are referred
to~\mycitet{Achterberg}{achterberg07} for a detailed description of the elements
of a state-of-the-art solution methodology for MILPs.

Binding together this collection of component algorithms is a control
mechanism that determines how to mix them effectively. The control mechanism
must make decisions about which implementation of each internal component to
employ (there are often multiple implementations to choose from), when to employ
them, and what amount of effort, relative to other components, should be
afforded.  This involves both auto-tuning to adjust strategies based on learned
properties of individual instances and controlling various trade-offs based on
user priorities and overall algorithmic strategy.

Subtle differences in algorithmic choices and control mechanisms may have a
dramatic effect, even if the effectiveness of these choices is not itself the
subject of investigation. This is what makes it particularly difficult to
rigorously judge the impact of a change in a single component within different
implementations or to compare different implementations overall.
There is no single approach to mitigating these issues, but awareness
is certainly the first step. What strategies are appropriate for overcoming
these difficulties must be determined on a case-by-case basis. 

\paragraph{Parallel Implementations.} Parallel implementations have all of the
same components mentioned above plus an additional control mechanism whose job
it is to distribute both tasks and associated data to the available
computational resources to enable parallel execution. This mechanism may either
be an integrated part of the overall implementation or may be completely
separated from the underlying sequential implementation. In the latter case, the
parallel control mechanisms are usually designed to work with any sequential
implementation whose external interface supports certain functionality required
for managing the data and task distribution process.  We refer to such a
separate control mechanism as a \emph{parallel framework}.

Parallel frameworks (and even integrated parallel control mechanisms) have their
own empirical behavior that can be evaluated independently of the underlying
solver implementation. However, isolating the effect of the parallelization
strategy from effects attributable to the properties of the underlying solver
implementation is difficult.
A given parallel control mechanism may work well with one solver implementation
and not with another. Further, it is primarily the \Scal that matters when
evaluating the scheme for parallelization, whereas it is primarily the baseline
\Perf that matters in evaluating the solver implementation. Teasing out the
effect of these various aspects of the overall implementation is difficult at
best, but is made especially so by the inherent trade-off between \Perf and
\Scal.

When the parallel control mechanism is tightly integrated with the underlying
sequential implementation, this may make it particularly difficult to isolate
the effect of its strategy on overall \Scal. On the other hand, it is easier to
observe the effects of parallelization strategies when using parallel frameworks
such as the \emph{Ubiquity Generator} (\ug) framework~\mycitep{UGweb} or the
CHiPPS/ALPS framework~\mycitep{xu09}.  Both \ug and CHiPPS consist of
collections of base classes that can be customized to parallelize any underlying
sequential implementation. \ug is designed to exploit the underlying sequential
solver as a black box, while CHiPPS requires more granular access. In both
cases, the modular implementation enables the use of the same underlying
sequential implementation with different parallel frameworks or the same
parallel framework with different underlying sequential implementations.  The
\ug framework, for example, has been customized to work with
\scip~\mycitep{scipopt320,shinano12,shinano13fiber},
\xpress~\mycitep{xpress,paraxpress}, and
\pipssbb~\mycitep{munguia16,Munguia2019}.

\subsection{Variability \label{sect:variability}}

One of the most challenging issues in assessing algorithms is that there are
many sources of variability that arise in measuring outcomes. This variability
can result in what appears to be anomalous
behavior~\mycitep{brui:async,Lai1984}. The sources of variability can be
divided roughly into two categories: \emph{Non-deterministic behavior} and
\emph{performance variability}. Non-deterministic behavior is the possibility
that an algorithm will follow different (but valid) execution paths at different
times---even when given precisely the same input---due to exogenous effects.
Performance variability, or algorithmic variability, is the possibility that an
algorithm will follow different execution paths given different but
\emph{mathematically equivalent} inputs. Performance variability can be
witnessed in conjunction with deterministic and non-deterministic behavior.

As a result of this variability, measures of effectiveness for an
implementation should be considered random variables and the outcome of
any \emph{single} computational experiment should be treated as one
observation. Repetition of experiments and the evaluation of results by
statistical methods are truly necessary, though either are rarely performed when
assessing the effectiveness of implementations. Space constraints prohibit the
inclusion here of a detailed discussion of proper methods for doing such
statistical analysis, but the paper of~\mycitet{Coffin and Saltzman}{CofSal00}
provides an excellent overview. In what follows, we discuss only the reasons why
such variability arises in the first place.

\subsubsection{Non-deterministic Behavior}

The possible non-deterministic behavior of implementations is a critically
important issue in empirical studies. Non-determinism is undesirable for
practical reasons in many settings. For example, customers of commercial
software vendors tend to distrust software that does not behave in the same
way when executed multiple times with the same input. Nevertheless,
determinism is sometimes hard to achieve. In particular, parallel
implementations are inherently non-deterministic unless specifically
implemented in a fashion designed to combat this. The reasons are related to
the order of execution of commands in a parallel computing environment.  For
example, different cores may operate at slightly different speeds and this can
result in sets of tasks being completed in different orders during two
executions of the same implementation. This non-determinism can trigger
cascading effects that eventually result in wildly different behavior from run
to run in some cases.

\begin{table}[b]
  \centering
  \small
  \begin{tabular}{l|rr}
    & Run time & Nodes \\
    \hline Run 1 & 620.834 & 161123 \\
    Run 2 & 560.324 & 159151 \\
    Run 3 & 578.838 & 165963 \\
    Run 4 & 652.507 & 178255 \\
  \end{tabular}
  \caption{The run time and number of created nodes for \symphony solving \emph{neos-1109824} using 4 cores over 4 different runs.}
  \label{tab:symphonyNondeterministic}
\end{table}

Table \ref{tab:symphonyNondeterministic} presents a simple example of
non-deterministic behavior of a parallel \bnb implementation. In this example,
\symphony is executed on four cores in order to solve \emph{neos-1109824} four
separate times. The run time and number of created nodes for each of the runs is
presented. From this small example, the largest deviation in run times is 14\%
where the largest deviation in the number of nodes is 11\%. This highlights a
real effect of non-determinism and the difficulty it introduces to the
assessment of algorithms.  For a more detailed example, the reader is referred
to the online supplement of \mycitet{Shinano et al.}{shinano13fiber}, which
presents the results of five repeated runs of FiberSCIP with different parameter
settings and solver configurations for all \miplib{}2010 benchmark instances.

It is possible to develop deterministic parallel implementations, but this
comes at the cost of reduced \Perf and \Scal due to synchronization steps that
introduce core idle time into the computation. On the other hand, it is
important to be aware that approaches to enforcing determinism do so by
ensuring that tasks get completed in the same order each time (at least for
all cases where the order matters). Nevertheless, this ordering should still
be considered arbitrary and the efficiency of the particular ordering enforced
may be much better or much worse than what would be observed with alternative
orderings.

\subsubsection{Performance Variability \label{sect:performanceVariability}}

\mycitet{Danna}{danna08} was the first to highlight the importance of
accounting for performance variability when assessing the effectiveness of
implementations. The variability can be of several types, some of which are
further described by~\mycitet{Lodi}{lodi13}. It is important to understand
that even when an implementation is sequential and deterministic in principle,
it may still exhibit performance variability. For example, permuting the rows
and columns of the constraint matrix in a linear optimization problem without
changing the problem itself may result in different behavior. This is because
the order in which the problem is read by the implementation can lead to
differences in procedures that are sequence dependent.  Some changes include the
way ties are broken, or choosing between alternative optimal solutions that are
produced when relaxations are solved.  These differences can be very large in
some cases.

An example of performance variability is presented in Table
\ref{tab:scipPermutation}.  In this example, \scip is used to solve
\emph{neos18} with 5 different random seeds for permuting the constraints of the
problem. It can be seen in Table \ref{tab:scipPermutation} that simply changing
the order of the constraints can have a significant effect on the \Perf of the
\bnb algorithm.

\begin{table}[h]
  \centering \small
  \begin{tabular}{l|rrr}
     & Run time & Nodes & LP Iterations \\
    \hline Seed 1 & 26.94 & 4154 & 280546 \\
    Seed 2 & 56.14 & 14446 & 653459 \\
    Seed 3 & 21.95 & 6033 & 201103 \\
    Seed 4 & 99.54 & 18032 & 1249410 \\
    Seed 5 & 30.56 & 4418 & 279840 \\
  \end{tabular}
  \caption{The run time, number of processed nodes and LP iterations for \scip
  solving \emph{neos18} using 5 different permutation seeds to permute the problem
  constraints.}
  \label{tab:scipPermutation}
\end{table}

While an implementation may \emph{appear} to be deterministic, the appearance of
determinism is deceiving.  The fact that different equivalent inputs can result
in very different \Perf means that the behavior is essentially non-deterministic
from the standpoint of a rigorous algorithmic analysis. Simply running the
implementation one time is not enough to understand its behavior. This kind of 
non-determinism is explained and exploited algorithmically in~\mycitet{Shinano
et al.}{shinano13fiber} and~\mycitet{Fischetti et al.}{fischetti15sample}.
The many sources of performance variability with respect to benchmarking are
discussed by~\mycitet{Koch et al.}{koch11}.
While no particular strategies are
proposed by~\mycitet{Koch et al.}{koch11} to address performance variability,
the discussion highlights the related issues and challenges.

\subsection{Impact of Test Sets \label{sect:instanceSelection}}

Test sets for the evaluation of \bnb implementations are abundant and also
highly influential. The most well-known and valued test sets for MILPs are
the \miplib series~\mycitep{achterberg06,bixby92,bixby98,koch11,miplib2017}
and the many test sets available for specific problem classes, e.g.,
TSP~\mycitep{tsplib}, QAP~\mycitep{qaplib}, and VRP with time
windows~\mycitep{solomon87}. These test sets have been developed with the aim
of providing an objective benchmark for the evaluation of different
implementations.

Although these test sets have generally proven to be useful to the research
community and helpful in aiding objective assessment, care must still be taken
when using these test sets. With respect to a single set of experiments, it is
important to determine whether a given test set is actually appropriate for the
experimental hypotheses in question. For example, it is not suitable to evaluate
a new branching rule using the complete \miplib{}2010 benchmark test set, since
a large proportion of the instances are solved in a small number of nodes (\scip
3.2.1 (with \soplex 2.2.1 as LP solver) solves 10\% of the instances in less
than 10 nodes).  Similarly, primal heuristics are best evaluated on problem
instances where it is \emph{difficult} to find the optimal solution.

More broadly, care must also be taken regarding the influence that publicly
available test set has on the research community agenda. The research
community should be striving not only to improve the effectiveness of
algorithms on a particular test set, but also across the broad range of
instances that the test set is meant to represent.

The general goal of selecting a test set is to select a subset of instances
that is ``representative'' of the instances in an entire class. The
effectiveness of an implementation, as assessed by the techniques described in
this paper, on the resulting benchmark computation, should be similar to what
one could expect in practical usage. In other words, summary measures, which
will be described in Section~\ref{sect:aggregationAndVisualisation}, should be
similar on the benchmark to what they would theoretically be across the entire
set of instances in the class. It is difficult to know for sure whether this
goal has been achieved, since for most classes of mathematical optimization
problems, we don't actually know what behavior is like across the entire set
of instances we're interested in. In general, we will likely have only seen
(and may only ever see) a very tiny subset of the possible instances to date.
The selection of instances for the MIPLIB 2017 benchmark set is a good example
of employing systematic methods to ensure the test set is
representative~\mycitep{miplib2017}.

\paragraph{Assessing \PERF.}

Test sets like \miplib provide a diverse set of instances with a wide range of
properties that are selected in an effort to allow a fair comparison of
different \mip solvers.  However, it is not always the case that test sets
appropriate for a particular class of problems will exist. When an appropriate
test set is not available and one needs to be gathered or constructed, it is
important to keep certain the following guidelines in mind.

It is well-known that, unless one has deep knowledge of the types of instances
that arise in a particular practical setting, randomly generated instances are
usually not representative (see~\mycitep{Joh02} for a related discussion of
this).  Rather, it is usually much better to gather instances that have actually
arisen in a particular application setting. Identifying particular properties of
instances that seem to have an impact on difficulty and ensuring that the
benchmark is diverse with respect to these properties is a good general
practice. Most importantly, the benchmark must avoid bias with regard to one
particular (type of) implementation. This is exceedingly difficult to achieve,
since the benchmark must be at least first designed prior to any testing having
taken place. For this reason, it is necessary to update benchmarks as more
knowledge of algorithms for a particular class is derived. Below are more
specific guidelines for constructing benchmarks specifically for the measurement
of \Perf and \Scal.

Performance variability is also an important issue that must be considered with
respect to instance selection. A factor not often considered is that the
instances of a chosen test set should not be particularly susceptible to large
changes in measured \Perf with only minor algorithmic changes. The report for
the \miplib~2010 test set comments directly on this issue~\mycitep{koch11}. In
particular, small test sets can be more prone to performance variability. As
such, it is important to use a test set that is diverse and contains enough
instances to reduce the negative effects of variability.

Although the following may be obvious, it should nonetheless be noted that the
\Perf of an implementation can not be assessed by looking at the results of a
single instance or even a small set of instances. After the release of
\miplib{2010} the overall geometric mean when measuring the \Perf of \cplex,
\gurobi and \xpress was nearly equal, while the \Perf on individual instances
varied by a factor of up to 1,500.

\paragraph{Assessing \SCAL.}

It is important to realize that properties of individual instances can limit the
\Scal that is possible to achieve, independent of a given implementation's
approach to parallelizing \bnb. In the context of \mip solvers, instances that
can be solved in a small number of nodes or for which the \lp relaxation in the
root node is extremely difficult to solve will not scale with any current
parallelization approach. For example, from the \miplib{}2010 benchmark test
set, there are 75, 63 and 60 instances, out of 86 instances, that require more
than 1000 nodes when solved using \symphony (Build Date: 24th January 2017),
\scip 3.2.1 (with \soplex 2.2.1) and a commercial solver respectively (as at
17th August 2017).  It is therefore important to select instances that are
suitable for evaluating parallel \bnb implementations.

\begin{itemize}
  
  \item Instances should produce a tree suitably large and broad enough that
    parallelization is both necessary and effective. Unfortunately, this
    property depends very much on the \Perf of the underlying sequential
    implementation. An instance may be suitable in this regard with respect to
    one implementation and not with respect to another. In addition, the size
    of the tree is not fixed and may vary based on random factors that induce
    performance variability.

  \item If one wants to use the traditional measure of speed-up, which is based
    on \wct, as a measure of \Perf, it is important that instances be solvable
    with one core (or at least a small number of cores). This will be used as a
    baseline for assessing the amount of overhead introduced by parallelization.
    Unfortunately, instances that can be solved in a reasonable amount of time
    on a single core may not be difficult enough with a large number of cores to
    be interesting and also not be suitable with respect to the first criteria.

\end{itemize}

\noindent This makes standard benchmark sets only partly useful to test parallel
\Scal, at least if we are assessing methods that parallelize the tree
search itself. Naturally, subnode parallelism could be employed in the case of
small trees, but as yet, this approach has not been vigorously pursued.

While the \miplib{}2010 benchmark test set is not ideal for assessing
\Scal, it is deemed suitable for the examples that we present here in this
paper.
A major reason for using the \miplib{}2010 benchmark test set is that there does
not currently exist a benchmark test set for assessing parallel \bnb
implementations. The creation of such a test set is an area of future research.

\subsection{Generalizability of Results}

One of the biggest pitfalls of assessment is the danger of over-generalizing
results and making conclusions that have little scientific basis. Given all of
the above challenges, it is critical to understand the limitations of any
computation with regard to its predictive power. As we typically have no
statistical basis for drawing conclusions and little knowledge of the true
properties of the entire set of instances in a given class, our ability to
generalize results from most experiments is extremely limited. At best, one can
only draw conclusions regarding effectiveness on the benchmark computation
itself and then only on the platform used in the experiment. It is important not
to over-state what one can conclude.

The comparison of scalability across parallel implementations is an area where
generalization is particularly difficult. In Section
\ref{sect:assessmentChallenges}, we discuss the difficulty in comparing parallel
implementations due to a lack of abstract frameworks. A particular difficult
arising from this situation is that most parallel frameworks are tightly linked
to a specific sequential implementation.  Thus, the \Perf and \Scal of parallel
implementations is highly dependent on the baseline \Perf of the underlying
sequential implementation.  Since it is not yet possible to control for baseline
sequential implementation \Perf when assessing \Scal, it is almost impossible to
compare different parallel implementations. As such, care must be taken when
making general claims about the \Scal of parallel control mechanisms for \bnb.

Finally, we should emphasize once again that the analysis of computation results
is largely a statistical one. Without repetition of experiments or some basis
for understanding the inherent variability of the results, one must acknowledge
that the results are at best only a rough estimate of reality. Although there
are cases in which one might reasonably expect the variability to be low, this
should be considered more the exception than the rule.

\subsection{Reproducibility and Verifiability}

Finally, we briefly mention the challenge of verifying results and ensuring
reproducibility. From the earliest days of scientific inquiry, one of the
fundamental principles of good science has been that scientific experiments
must be verifiable (in the case of software, this means verifying correctness
of the implementation) and reproducible. The fact that these principles should
also be applied in empirical assessment of algorithms is noted
by~\mycitet{Johnson}{Joh02}, among others. Unfortunately, few experimental
results in the realm of computation satisfy these two minimal requirements of
scientific studies. The challenge of how to ensure verifiability and
reproducilibity is a very tricky one and the conversation around this topic is
only just getting started. In most cases, making the source code used for the
experiments available to others is the only way to truly ensure both
verifiability and reproducibility. Providing executable code without source
may allow reproducibility, but prevents verification of correctness. Although
we do not have any answers and understand the realities surrounding this
important issue, we would be remiss if we did not make it part of the
discussion surrounding challenges of empirical research.

\section{Measures of Effectiveness \label{sect:comparisonMetrics}}

The framework of Section~\ref{sect:performanceAndScalability} introduced the
concept that the effectiveness of implementations can be thought of as the
efficacy with which an implementation consumes computational resources in both
the sequential and parallel cases. Implementations can be assessed with
respect to their baseline \Perf or their parallel \Scal, with the main
features in the analysis being how one designs the experiments and how one
decides what resource levels to vary and which ones should be held constant. In the
case of \Perf analysis, the resource that is typically measured is CPU
time, with all other resources fixed, including the number of cores available
to the computation. In a typical (strong) \Scal analysis, on the other hand,
one tries to measure the trade-off between \wct and
the number of cores.

Many challenges associated with the assessment of \Perf and \Scal are common
to both types of analysis. Although both types of analysis can be based on
measures of resource consumption in principle, direct measurement can be
difficult and proxies are often used. The proxies that are appropriate for
assessment are different in each case and the methods of analysis may also be
different. In what follows, we first review the assessment of \Perf and then
that of \Scal.

\subsection{\PERF \label{sect:performanceMetrics}}

\Bnb implementations typically yield a number of convenient measures for use
in assessing \Perf that are readily available from the solution output log.
Below, we review properties of specific measures typically used in assessment.

\subsubsection{Measuring Resource Consumption \label{sect:measuringResources}}

Directly measuring the resources needed to achieve a given termination
criteria is the most effective way to determine an implementation's \Perf.
As we have mentioned, the resource that is typically measured in assessing the
efficiency of an implementation is time (CPU or wall clock), though one
could easily measure the consumption of any other resource of interest when
appropriate. It should be noted that time may itself be a proxy for what we
truly care about in some cases. In practical experimental settings,
interference from other experiments running simultaneously or from the
operating system itself may cause variability in the time measurement that
must be accounted for. Measuring work (atomic operations), as discussed below,
may be an alternative in cases where this a serious problem.
Below, we discuss the termination criteria typically used. 

\paragraph{Time to Optimality.}

The most common measures of \Perf is the \emph{time to provable optimality}.
This measure is easily understood and is a direct measure of what is often
desired by users, as well as what a \bnb implementation is truly designed to
do---determine an optimal solution and construct a corresponding proof of
optimality as quickly as possible. While this is perhaps the most common way of
measuring efficiency, it has one big drawback---it does not provide a way of
handling assessment of instances whose computation doesn't complete within the
specified time limit.

\paragraph{Time to Fixed Gap.}

In many practical applications, a provably optimal solution may not be needed.
Rather, a solution of a (provably) acceptable quality is sufficient. For the
purpose of assessing the quality of the best known solution, we may use various
measures based on the difference between the primal and dual bounds, the
so-called \emph{optimality gap}. The value given by simply subtracting the dual
bound from the primal bound is referred to as the \emph{absolute optimality gap}
or \emph{absolute gap}. When the primal and dual bounds are equal, the absolute
gap is zero and the optimal solution value has been found. Alternatively, a
scale invariant version of the gap may be desired. This is given by computing a
normalized difference between the primal and dual bounds, the so-called the
\emph{relative primal-dual optimality gap} (or just \emph{primal-dual gap}).
This gap can be computed by
\begin{equation}
  \gamma(Z^{p}, Z^{d}) = \left\{
    \begin{aligned}
      &0, && \text{if } |Z^{p}| = |Z^{d}| = 0, \\
      &\frac{|Z^{p} - Z^{d}|}{\max\{|Z^{p}|,|Z^{d}|\}}, && \text{if } Z^{p} \times Z^{d} \geq 0, \\
      &1, && \text{otherwise.}
    \end{aligned}
    \right.
    \label{eqn:primaldualgap}
\end{equation}
where $Z^{p}$ and $Z^{d}$ are the primal and dual bounds on the optimal
solution value, respectively. The \emph{relative primal gap} and
the \emph{relative dual gap} can be computed by replacing $Z^{d}$ or $Z^{p}$
with the optimal solution value or the best known primal or dual bounds,
respectively. From here on, the general term ``gap'' will refer to any of the
above variants of the relative optimality gap. Note that
in \eqref{eqn:primaldualgap}, we have set the denominator of the second case to
the maximum of the absolute primal and dual bounds. This has been done so that
the gap is always between 0 and 1. Other definitions of the gap set the
denominator to the minimum, instead of the maximum, or even set the
denominator to an appropriate constant value.

The time to reach a given gap is a measure of an implementation's
ability to both produce a solution of a desired quality and \emph{provide a
proof of its quality}. Although similar to time to optimality, time to fixed
gap is a more suitable measure for practical applications where an acceptable
solution quality can be defined.

\paragraph{Time to First Solution.}

A third alternative termination criteria, often acceptable in practical
applications, is the discovery of any one feasible solution. Although no proof
of quality need be provided in this case, it may still be an appropriate
criteria when feasible solutions are difficult to find or when optimality is
not as important as speed.
The time to first solution primarily measures the effectiveness of the \bnb
implementation's primal heuristics.
Obviously, since the \Perf according to this measure relies mainly on the
primal heuristics, large parts of the implementation are neglected. In
particular, the ability to improve the lower bound (the dual side of the
algorithm) is ignored. Hence, this measure only provides a very narrow view of
an implementation's \Perf. It is also likely that this is a measure subject to a
much higher degree of variability than some others, as a greater degree of
``luck'' is involved in locating a solution quickly.

\subsubsection{Measuring Work Performed \label{sect:measuringWork}}

Measures of work count the number of atomic operations required to complete a
given computation. In many cases, it can be expected that the atomic
operations being measured take a constant (and predictable) amount of time on
average, so that the work preformed and the time it takes will be linearly
related. Note that there is a difference between what is considered ``atomic''
with respect to the type of measurement considered in the section below and
what is considered ``atomic'' when it comes to the measurement of CPU time. In
the latter case, what is considered atomic could vary based on the particular
operating system and compiler used, as well as details of the hardware itself,
whereas the atomic operations considered below are much higher-level and depend
only on details of the algorithm implementation. This can make the measures of
work described here more appropriate for the purposes of comparison in some
cases, but this depends on the type of experiment being performed and what
hypothesis is being tested. 

The assumption that atomic operations take a predictable amount of time on
average may not hold in some important cases. In particular, when the number of
available cores is scaled, this will generally increase the time per atomic
operation due to the introduction of overhead, as described in
Section~\ref{sect:scalabilityMetrics}. Scaling the amount of memory available
can also affect the time per atomic operation.

We would like to highlight that care must be taken when using the measurement of
work as a proxy for time. In the remainder of this section, we review the most
typical measures of work employed in the context of \bnb.

\paragraph{Number of Nodes.}

For a \bnb implementation that is fully deterministic, the number of nodes
enumerated to solve a problem to optimality is a measure of work performed by an
implementation. This measure of work can sometimes be used as a reliable proxy
for the time required to solve an instance to optimality. However, this proxy
relies on the assumption that the processing time for each node is identical.
Typically, this is not true, even for deterministic sequential implementations,
since different component algorithms (see Section
\ref{sect:assessmentChallenges}) will be executed at different nodes throughout
the tree. This assumption is particularly tenuous when comparing different
implementations or parameter settings or running on a different number of cores.
For example, generating cuts at every node in the tree will increase the time
required to process a node, but is expected to reduce the number of nodes that
need to be enumerated.  When using the number of nodes as a measure of \Perf, it
is important to understand what algorithms are executed during node processing
in each implementation.

The number of nodes \emph{is} an important measure when considering the effect
of changes in certain parameter settings or when considering the effect of
scaling the number of cores. As described above, putting more effort into node
processing may reduce the number of nodes produced at the cost of increased
time. Even with a single implementation, changes to parameter settings can
have a big effect on the size of the search tree for this reason. More
importantly, the number of nodes produced can generally be expected to increase
when additional cores are used in parallel computation and this increase is one
component of the overhead, discussed later in
Section~\ref{sect:scalabilityMetrics}.

\paragraph{Number of Bounding Problems Solved.}

During the processing of each node, it is typical to solve a sequence of
bounding problems. This is the case, for example, when the bound is computed
by solving a relaxation that can be strengthened once the result of the
computation is known. By iteratively strengthening the bounding problem, it
may be possible to decrease the number of nodes that must be processed
overall, but this comes at the cost of increasing the amount of work done per
node. Thus, one could argue that the number of nodes processed does not give a
complete picture of the effectiveness of a given (variant of) an
implementation, making the processing of a single node a less-than-ideal
atomic unit of work. Since the number of bounding problems solved is a (rough)
measure of the amount of work performed in processing each node, one could
argue that the total number of boundings problems solved is a more fine-grained
measure of work.  For example, if an aggressive cut generation strategy is
employed, this would result in more bounding problems (LPs) being solved at each
node and fewer nodes processed. The number of bounding problems solved in total,
however, could go up or down.  Ideally, the combination of the number of nodes
processed and an assessment of the number of bounding problems solved per node
should give a more complete picture of the overall effectiveness of a given
implementation.

\paragraph{Iteration Count.}

In most cases, the algorithm used to solve the bounding problem is iterative
and the \Perf of \emph{this} algorithm can itself affect the effectiveness of
the overall \bnb algorithm. For example, for \lp bounding problems a number of
pivots must be performed to find the optimal solution. Similarly, for interior
point algorithms, solution of the bounding problem typically requires a number
of function evaluations. The total number of iterations performed in solving
bounding problems over all nodes in the \bnb tree is called the iteration
count and can provide an even finer-grained measure of work than the number of
bounding problems solved.

\subsubsection{Measuring Progress \label{sect:measuringProgress}}

As we mentioned earlier, measures of progress are important in assessing
effectiveness when computations cannot be fully completed within a given time
limit. Rather than measuring the resources needed to complete a computation,
measures of progress are used to assess the amount of computation that can be
done with a fixed resource configuration. By its very nature, what is meant by
``progress'' in \bnb is not easy to define. On the one hand, one can define it
in terms of the fraction of ``useful computation'' that has been performed. On
the other hand, the solution of NP-hard problems inherently involves a certain
amount of backtracking, which results in little perceivable progress, other
than eliminating one particular dead end. Although the exploration of dead
ends can be avoided in principle with some guided ``luck,'' there is no
reliable way to ensure this in general (provided P $\not=$ NP) and thus it is
difficult to imagine that any rigorous way can be developed to determine the
fraction of ``useful computation'' that has been completed.

Despite the inherent difficulty in measuring the progress of a computation, it
is nevertheless essential, in many scenarios, to report some measure of
progress. A particular use of a measure of progress is to inform users of a
software package about what progress has been made during a long calculation.

In the context of assessment, a reliable measure of progress would provide a way
of assessing effectiveness without performing the full computation.
Interpreted in another way, by assessing implementations using a measure of
progress, we hope to be able to determine the \emph{rate} of progress for each
possible resource configuration. If we had an accurate estimate of how long it
takes to do some fixed fraction of the computation, this would be enough to
enable a rigorous assessment. Alas, this scheme hinges critically on having a
reliable measure of progress and such a measure does not currently exist. The
measure of progress most widely used at the moment, the optimality gap, is
described below, along with two alternative, the gap integral and tree size
estimates.

\paragraph{Gap.}

\begin{figure}[bp]
  \begin{center}
    \subfigure[Implementation 1 - Bounds]{
      \includegraphics[width=0.31\textwidth]{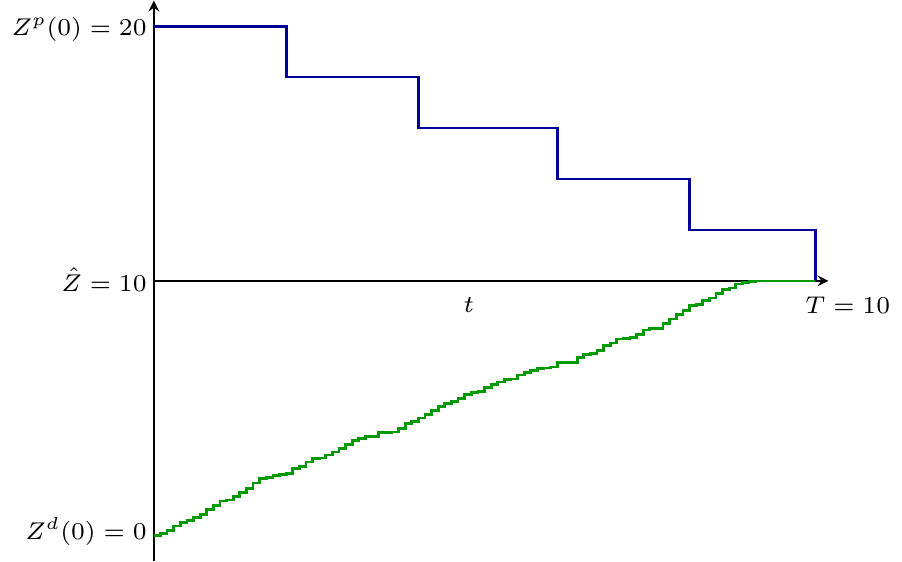}
      \label{fig:boundEvolution1} }
    \subfigure[Implementation 2 - Bounds]{
      \includegraphics[width=0.31\textwidth]{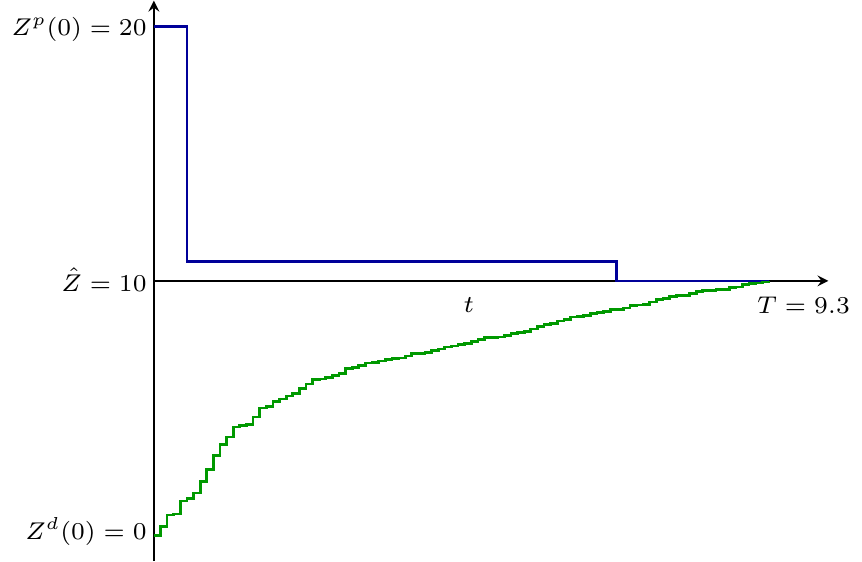}
      \label{fig:boundEvolution2} }
    \subfigure[Implementation 3 - Bounds]{
      \includegraphics[width=0.31\textwidth]{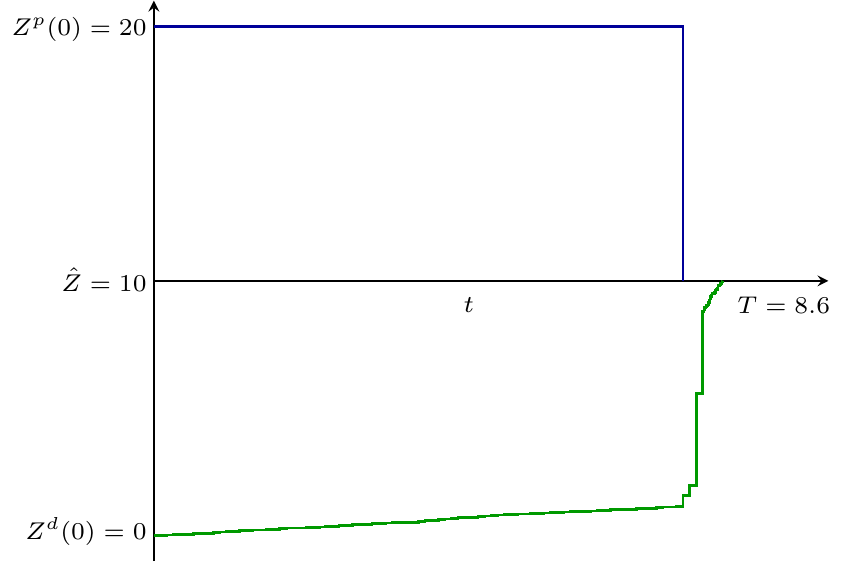}
      \label{fig:boundEvolution3} }
    \caption{Graphs exemplifying the evolution of upper and lower bounds.}
    \label{fig:boundEvolution}
  \end{center}
  \vspace{-5mm}
\end{figure}
The most often cited measure of progress is some variant of the optimality gap
described in Section~\ref{sect:performanceMetrics} of which the primal-dual gap
is the most commonly used. This gap seems natural as a measure of progress,
since it decreases monotonically throughout the computation and is 0 when the
computation is completed. Unfortunately, though, the evolution of the gap may be
far from smooth and it leaves much to be desired as a rigorous measure of
progress. To illustrate, Figure \ref{fig:boundEvolution} shows three examples of
the evolution of the primal and dual bounds during the solution process of three
different implementations when solving an MILP. In this figure, the blue and
green curves are the functions $Z^p(t)$ and $Z^d(t)$ whose values are the primal
and dual bounds, respectively, at time $t$.  
\noindent One feature highlighted is that the biggest changes typically come
from improvements in the upper bound, while evolution of the lower is
generally, though not always, smoother. Figure \ref{fig:boundEvolution1}
exhibits a constant, regular improvement in the upper and lower bounds,
leading to a relatively smooth (though not linear!) evolution of the gap. This
is not the typical case.
More often, the gap evolution in a \bnb implementation is characterized by
large changes in the upper bound and long periods without any improvement, as
demonstrated in Figures \ref{fig:boundEvolution2} and \ref{fig:boundEvolution3}.
In those two cases, after 5 seconds of computation, the value of the gap would
suggest that the former implementation outperforms the latter. However, after
the full computation, the total run time for the former implementation is
greater than that for the latter. This small example highlights the difficulty
in using gap to measure progress.

\begin{figure}[tbp]
  \begin{center}
    \subfigure[Implementation 1 - Gap]{
      \includegraphics[width=0.31\textwidth]{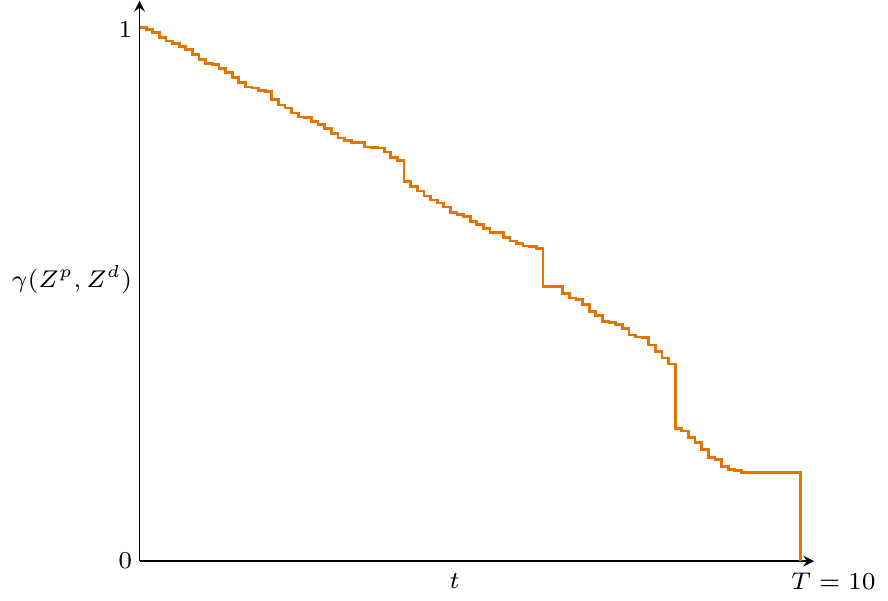}
      \label{fig:gapFunction1} }
    \subfigure[Implementation 2 - Gap]{
      \includegraphics[width=0.31\textwidth]{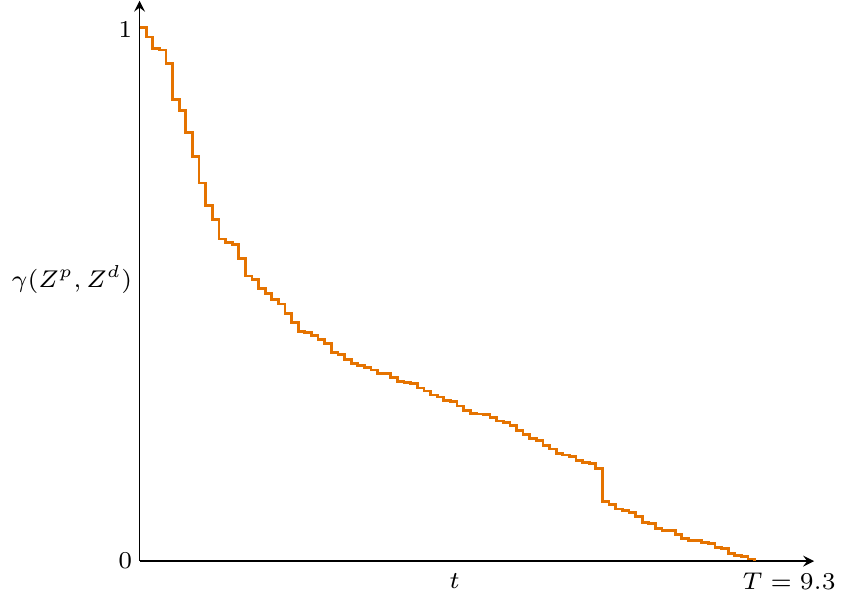}
      \label{fig:gapFunction2} }
    \subfigure[Implementation 3 - Gap]{
      \includegraphics[width=0.31\textwidth]{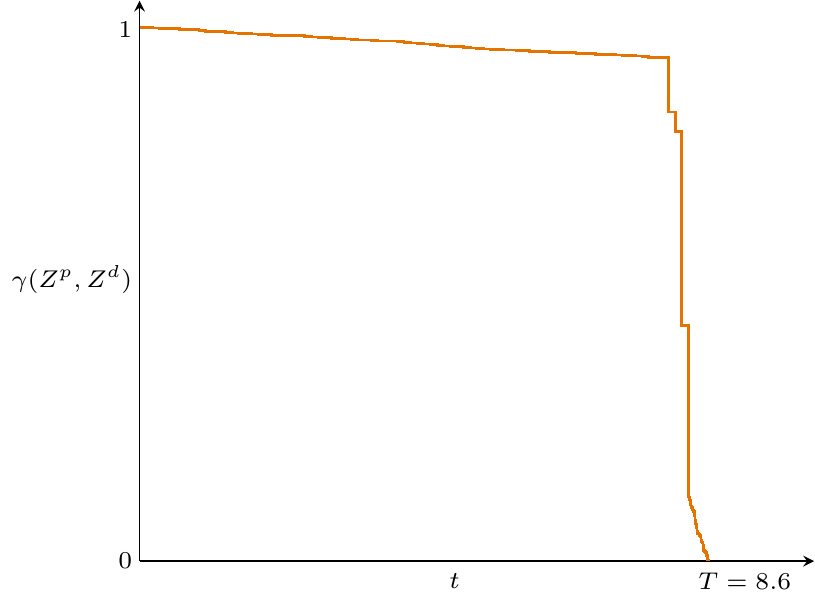}
      \label{fig:gapFunction3} }
    \caption{The gap functions corresponding to the graphs in Figure~\ref{fig:boundEvolution}}
    \label{fig:gapFunctions}
  \end{center}
  \vspace{-5mm}
\end{figure}

Figure~\ref{fig:gapFunctions} shows the graph of the \emph{gap
function} whose value at time $t$ is $\gamma(Z^{p}(t), Z^{d}(t))$, where
$\gamma$ is as defined in~\eqref{eqn:primaldualgap}.  
The shape of the gap function reflects the evolution of the primal and dual
bounds, though some information is clearly lost. Based on these graphs, it is
easy to see why the gap function may not have much power as a measure of
progress. Predicting progress using this function would require knowing the
general shape of the gap function ahead of time and this is exactly what we
generally do \emph{not} know. 

Despite the fact that the gap has limited predictive power, it may still be
used effectively as a means of comparing the progress made by two different
algorithms in solving the same problem with the same resources, although even
this usage must be approached with some caution. Large differences in gap may
not be as significant as they appear. In
Section~\ref{sect:aggregationAndVisualisation}, we describe a method of
incorporating gap into an overall visual summary of results that can allow it
to be used more meaningfully.

\paragraph{Gap Integrals.}

Gap integrals are an alternative to the more classical measures of \Perf and
also provide an alternative to the gap itself as a measure of progress. Rather
than simply measuring the gap achieved for a fixed time limit, the gap
integral is computed from a time series that captures the evolution of the gap
over a time-limited computation. Roughly speaking, the gap integral measures
the ``average gap'' over a given time interval. While it can be viewed as a kind
of measure of \Perf, we would like to suggest here that gap integrals may also
be useful as a measure of progress.

The concept of a gap integral was first proposed
by~\mycitet{Berthold}{berthold13}. The specific type of gap integral first
proposed is known as the \emph{primal integral} because it is based on the
evolution of the primal gap. Corresponding gap integrals can be obtained based
on other notions of optimality gap. We focus here on the
related \emph{primal-dual integral} (PDI), which is based on the primal-dual
gap given by~\eqref{eqn:primaldualgap}.

To formally define the PDI, we interpret the primal and dual bounds
in~\eqref{eqn:primaldualgap} as functions of time $t$, as above. Then, the primal-dual
integral is given by
\begin{equation*}
  PDI = \int^{T}_{0}\gamma(Z^{p}(t), Z^{d}(t))\text{d}t,
\end{equation*}
where $T$ is either the wall clock/CPU time needed to solve the instance or the
time limit (if the instance is unsolved). Since we usually only observe discrete
samples of $Z^p$ and $Z^d$, the primal-dual integral is typically approximated
by the more practical formula
\begin{equation*}
  PDI = \sum^{\mathcal{I}}_{i = 1}\gamma(Z^{p}(t_{i}),
  Z^{d}(t_{i}))\cdot(t_{i} - t_{i - 1}), 
\end{equation*}
where $t_{i} \in [0, T]$ and $i \in 1, \ldots, I$ are the discrete times at
which the primal or dual bounds are updated, with $t_{0} = 0$ and $t_{I} = T$.
Thus, the PDI is a number between zero and $T$, with numbers closer to zero
indicating a more effective implementation.

The use of gap integrals as a measure of effectiveness was originally motivated
by the observation that two different implementations that discover the same
quality of solution, but at different times in the evolution of the algorithm,
should not be judged equally in terms of effectiveness. Time to provable
optimality (or any other termination criteria) is agnostic to how the gap
actually evolved during the computation, but a gap integral rewards an algorithm
for making quick progress in reducing the gap in the beginning of the
computation, even if the eventual time to optimality is no better.
In many applications, it can be argued that the implementation that finds a
high-quality solution first is the better performer, regardless of eventual time
to optimality.

When one analyzes the evolution of the gap over the solution process,
differences in \Perf that are not made evident by other measures can be
observed. Consider the illustrative examples of gap evolution presented in
Figure \ref{fig:gapFunctions}. All three report a similar time to optimality,
but they all display very different progress in the gap over time. Classical
measures of \Perf would lead us to conclude that the third algorithm is
``best.'' However, the gap integral leads to a different (and equally valid)
conclusion. While the third algorithm had difficulty both in finding a good
initial feasible solutions and in improving the dual bound, the second algorithm
exhibited a much better performance from the primal heuristics and exhibits the
lowest PDI of the three. Although the gap integral produces a single summary
statistic, this number can be viewed as a summarization of a time series of data
about how the algorithm progressed.

\paragraph{Tree Size Estimation.} 

Methods of estimating tree size attempt to predict the final size of the tree
using data gathered during the algorithm's execution up to a given point in
time. Recent examples of online tree search estimation have been presented
by \mycitet{Belov et al.}{belov17} and
\mycitet{Anderson et al.}{anderson18}. If the final size of the tree could be
accurately predicted, then such a method would in turn provide an accurate
measure of what fraction of total work was completed as of a given time limit.
As shown in \mycitet{Anderson et al.}{anderson18}, it is possible to relatively
accurately estimate the final size of the tree at intermediate points during the
solution process. Having such an estimate would, in turn, help to derive a
reliable measure of progress: since the number of processed nodes can be
compared to the estimated total size. While current approaches have proven
valuable for triggering specific algorithmic features, such as solving restarts,
these efforts are unfortunately not accurate enough for use as a measure of
progress. Further, there is little reason to believe that such a method will be
discovered that will be accurate enough given currently accepted conjectures
about the complexity of solving mathematical programs (i.e., that $P \not= NP$).

\subsection{\SCAL \label{sect:scalabilityMetrics}}

In this section, we discuss measures for assessing strong \Scal in the
traditional way. As described in Section \ref{sect:scalability}, \Scal is a
measure of the trade-off between different resources or, alternatively, a
measure of the change in efficiency that arises when using different bundles of
resources to solve a problem. Most commonly, this trade-off is between the \wct
needed for solving an instance and the number of available cores. As such, many
of the measures used for assessing \Perf can also be used to analyze \Scal,
albeit in a different way.

The goal of \Scal analysis is to measure how efficiently one resource
(typically cores) can be used to save another resource (typically, time).
Ideally, this trade-off has a linear relationship---doubling the number of
cores cuts the required time in half. In practice, this situation is rarely
achieved because parallelization introduces additional resource consumption,
commonly described as \emph{overhead}.
\begin{definition}
\emph{Parallel overhead} is resource consumption that is only incurred due to
the parallelization of an implementation.
\end{definition}
\noindent Many of the sources of overhead are discussed in detail
in~\mycitet{Koch et al.}{koch12} and~\mycitet{Ralphs et al.}{RalShiBerKoc18}.
Here we will only provide a summary of sources and types of overhead. The most
obvious type of overhead is \emph{communication overhead}, which is the work
done in order to transfer data from one location to another. A second major
source is \emph{idle time}. This occurs when computation on one or more cores
is waiting on data that will only be made available after a specific
computation on other cores has completed. Overhead, observed as waiting time,
in parallel implementations can be introduced for multiple reasons.
\begin{itemize}
\item In distributed computations, data currently stored remotely might need
to be transferred to local storage for processing.
\item In shared-memory computation, simultaneous access to the same
block of data may be restricted due to a memory \emph{lock} to prevent memory
corruption and the introduction of race conditions.
\item It may be required that a certain remote computation is completed before
proceeding with the current local computation.
\item Finally, in the start or end of a computation, the available work
may simply not be granular enough or it may not be efficient to divide the
computation among all available cores, leaving some of them idle. This idle
time is known as \emph{ramp-up} and \emph{ramp-down} time.
\end{itemize}
Each of the sources described above commonly arise from the parallelization of
any algorithm, not only \bnb. A third major source of overhead that
arises specifically from the parallelization of \bnb implementations is
generally referred to as \emph{redundant work}. Roughly, this is additional work
that could have been avoided with better global knowledge of the state of the
computation.

Overhead can be measured either directly or indirectly. Both methods of
measurement are challenging for a variety of reasons. Direct measurement
involves explicitly capturing data on the contributions of various sources of
overhead, such as idle time or time spent in communication (see
Section \ref{sect:measuringOverhead} for more on these). The main challenges
of direct measurement of overhead are purely technical. For example, idle
time, discussed above, can occur for many different reasons, some of which are
directly attributable to parallelization and some of which are not. Some
sources of idle time are easy to measure and identify (time spent waiting for
data to arrive from another location in a distributed memory environment),
while others are not (time spent waiting for data to be fetched from shared
memory). Although many parallel \bnb implementations do measure and report
certain sources of idle time and other overhead, the various types of overhead
might be captured and reported differently.

Alternatively, indirect assessment of overhead involves observing the change
in either the \Perf, the amount of work performed, or the amount of progress
achieved for a given computation and assuming that any increase is due to
overhead. This is generally a more reliable and all-inclusive method of
measuring the overall level of overhead, but does not help in identifying the
sources of the overhead. Thus, the methods for reducing overhead remain unclear.

The assessment of \Scal is primarily about measuring the amount of overhead
and determining how it changes as more resources are added. Thus, \Scal is
somewhat independent of the baseline \Perf of the parallel implementation,
as we illustrated earlier. This introduces a difficulty in discussing and
comparing the \Scal of multiple implementations, since an implementation
exhibiting good \Scal may also exhibit poor \Perf, as shown in Table
\ref{tab:alpsKnapsack}, and vice versa.  Many of the measures presented in the
following sections are used to evaluate \Scal while being agnostic to the \Perf
of the implementations. While these measures may be efficacious for the
assessment of \Scal, one must also consider whether the reported measure can
still be useful when also accounting for \Perf. We don't propose a solution to
addressing this issue here, but mention that it is a focus of current research.

There are many different assessment measures that can be used to indirectly
measure overhead. These measures and how they are used for assessing overhead
is discussed in
Sections \ref{sect:changesEfficiency}--\ref{sect:changesProgress}. To measure
overhead indirectly, we must recognize the inherent variability in the path
taken by the computation. In almost any sophisticated algorithm for solving an
NP-hard problem, small perturbations in the path (due to differences in
tie-breaking, etc.) that arise at early stages of the algorithm could have
unexpected large effects, as discussed in
Section~\ref{sect:performanceVariability}. For any single computation, changes
in total resource consumption can fluctuate only due to effects from
performance variability. We must endeavor not to conflate these fluctuations
with changes in the resource consumption due to the introduction of true
overhead. A method used to alleviate this issue is to perform repeated
experiments over large test sets to draw statistically valid conclusions from
the results.

\subsubsection{Measuring Overhead Directly \label{sect:measuringOverhead}}

Direct measurement of overhead can be a difficult and work-intensive task.
However, the direct measurement of overhead for \Scal analysis has the advantage
of illuminating the precise reasons why an implementation is or is not scalable.
The types of overhead within a parallel implementation that can be directly
measured are as follows.

\paragraph{Ramp-up and Ramp-down Time.} Ramp-up and ramp-down time can generally
be measured rather easily. Although it must be pointed out that these concepts
can be defined in a number of related ways. One definition for ramp-up time for
an individual core is the time from the beginning of a computation until that
core is actively performing computation. The overall ramp-up time is then the
sum of the ramp-up times for individual cores. Ramp-down time can be defined
similarly as the elapsed time from the last computation done by a given core and
the end of the full computation. Alternatively, the total ramp-up time can be
defined as the time from the start of computation until the first time all cores
are active. Also, the ramp-down time could be defined as the time from the end
of the computation back to the last time all cores were active.  Regardless of
the definition, the ramp-up and ramp-down times are easily captured and
generally reported. Because of the differences in the definitions, one must take
care when comparing the ramp-up and ramp-down times reported by different
implementations.

\paragraph{Communication Time.} Most work related to communication can also be
effectively measured by simply recording the time spent in relevant parts of the
implementations code. In distributed parallel implementations, subroutines
dedicated to the job of packing and unpacking data to be sent and/or received
from remote locations are usually modularized and time spent in those
subroutines can be directly measured using a profiler.  

\paragraph{Idle Time.} Idle time is generally more difficult to measure and its
causes are more difficult to identify. Especially in the case of shared-memory
parallelism, where the locking of data and atomic process are necessary to avoid
memory corruption and data races. It may be possible to measure the idle time
and attribute it directly to a cause in some cases. For example, in distributed
parallel implementations, the time spent executing a blocking receive call is an
explicit measure of time spent waiting to receive data from a remote location.
Given the difficulties of measuring idle time, proxies based upon the solving
statistics of a \bnb implementation are typically used.

\paragraph{Node Throughput.} Node throughput is a measure of the overall
rate at which work is being performed and can be used as a measure of the
additional idle time introduced by the parallelization. Throughput simply
measures the number of nodes processed per second, or equivalently, the
average time it takes to process a single node. When an implementation is
parallelized, the processing of a node is expected to be the same regardless
of the number of cores used. This is because frameworks for parallelization
of \bnb typically only parallelize the tree search, so there should be no
computational difference in the processing of a single node when the algorithm
is parallelized. Any decrease in the node throughput can suggest that
additional work is being performed as a result of the parallelization.

Since the node throughput is an average measure, there are a number of issues
that must be considered when assessing \Scal. First, as explained above, it is
not always true that the time to process each node is identical in the
sequential and parallel implementations. A particularly poignant example are the
implementations of ParaSCIP~\mycitep{shinano12} and
FiberSCIP~\mycitep{shinano13fiber} that both perform presolving when a node is
transferred to another solver, which is typically running on a separate core.
This additional presolving step has been observed empirically to improve the
\Perf of ParaSCIP and FiberSCIP.  Since this involves additional work per node,
a decrease in the node throughput when increasing the number of cores could be
observed, incorrectly suggesting parallel overhead, when in reality, this is
because parallelization has introduced an algorithmic change. More
fine-grained parallelism may also result in more tasks being performed in
parallel during the processing of a node. While a node is expected to be
processed faster in parallel, as compared to a sequential implementation,
computational resources are diverted from the tree search to perform
alternative tasks. To the best of the authors' knowledge, there has been
little investigation into the impact of subnode parallelization to determine
its impact on the overall node throughput.

\subsubsection{Measuring Changes in \PERF \label{sect:changesEfficiency}}

The most common way of measuring scalability is to (indirectly) measure the
total overhead by measuring the change in measured efficiency that occurs when
utilizing additional cores for the computation. This is typically done by
comparing the actual wall clock time required for a given computation on $N$
cores (denoted $T_N$) to the baseline of ``perfect scaling,'' as described in
Section~\ref{sect:scalability}. This ideal baseline is given by $T_1/N$, where
$T_1$ is the sequential run time (or parallel run time on a single core). The
difference between the actual and ideal scaling is interpreted as overhead.

To make the assessment easier, statistics that are independent of the actual
running time can be computed by scaling in two different ways.
The \emph{speed-up} of a parallel implementation utilizing $N$ cores is $$S_N
:= T_1 / T_N,$$ so that the value of $S_N$ represents the factor by which the
wall clock running time of the implementation changes when more cores were
added. When the scaling of a parallel implementation is perfect, the speed-up
is $S_N = N$ (this is also called \emph{linear speed-up}). As discussed
earlier, it is typical to observe that $S_N < N$, which indicates there is
some loss of efficiency (increase in overhead) when employing $N$ cores. The
parallel efficiency $$E_N = S_N/N$$ is a measure closely related to speed-up
that is further scaled to obtain a number normally between zero and one,
making the statistic also independent of the number of cores. A parallel
efficiency of one represents perfect scaling and the value $1 - E_N$
represents the fraction of the total computation time that was attributable to
overhead.

The speed-up or parallel efficiency of a parallel implementation are commonly
used and easily interpreted statistics for assessing \Scal that are used
throughout the parallel computing community. However, these measure have a
number of weaknesses when employed in the assessment of parallel \bnb
implementations. One of the major difficulties of this approach, the existence
of performance variability, was described earlier in
Section~\ref{sect:performanceVariability}. Due to the effects of this
variability, it is highly probable that the execution paths of the sequential
and parallel implementations will be completely different. To exacerbate the
situation, parallel \bnb implementations are typically also non-deterministic,
since techniques for enforcing determinism introduce additional overhead. As
such, it is possible that two runs of the same parallel implementation with
the same number of cores will exhibit vastly different resource consumption.
Consequently, it is difficult to determine whether the loss of efficiency
should be attributed to the parallel implementation or a less performant
execution path. These challenges can be overcome with proper repetition and
statistical analysis of experimental data, which is vastly more important in
the parallel setting then in the sequential one.

A further limitation arising from the use of parallel efficiency as a measure
of \Scal is that it is typically based on the time to optimality (or some
other termination criteria). As such, this measure can only be used with
instances in which the sequential and parallel implementations can both find
the optimal solution. Hence, the instance selection plays a major role in the
observed \Scal. Because the \Perf of different parallel implementations can be
vastly different, it is very difficult to find a common test set that will
make this measure meaningful.

\subsubsection{Measuring Changes in Work Performed \label{sect:changesWork}}

Many of the measures of work presented in Section \ref{sect:measuringWork} can
also be used as proxies to assess certain aspects of the \Scal of parallel
implementations by comparing the total work done in executing the parallel
implementation to the work done in the sequential implementation. The most
common measure of work used for this purpose is the number of nodes processed.
Changes in the number of nodes processed are mainly due to two phenomena.
First, the order in which nodes are being searched is different, even when the
search strategy itself is the same. Since multiple subtrees are being explored
simultaneously, some nodes will inevitably be discovered relatively earlier in
the parallel implementation than in the sequential implementation. This can
lead to the processing of nodes in parallel that would not have been processed
in the sequential algorithm because of differences in the timing of discovery
of primal solutions. 

A second, though much less impactful, reason is the lack of accurate
information about bounds discovered at remote compute cores for reasons
described below. In deterministic parallel implementations, locally generated
bound information is usually broadcast only periodically, at specified
communication synchronization points. Non-deterministic \bnb implementations
broadcast bound information on a more continuous basis, but delays are still
inevitable. In both cases, there will inevitably be periods of time when the
bound information available at a given node is outdated. As a result, a
subtree search performed locally could end up processing nodes that have a
dual bound greater than the best global known primal bound. This phenomenon
causes an increase in the number of processed nodes for the parallel
implementation compared to the underlying sequential implementation. The
reverse phenomenon can also be observed in non-deterministic parallel
implementations. Since many nodes are processed in parallel and the improved
bounds are constantly communicated, it is possible that a node that improves
the primal bound is found earlier than in the sequential
implementation~\mycitep{Lai1984}. This would result in more nodes being
discarded earlier and would reduce the total number of nodes processed.

The obvious challenges in using measures of work to assess \Scal are similar
to those described at the end of the previous section, primarily due to the
inherently high degree of variability in the measurements. As such, one must
perform testing with due care and attention to proper design of the
experiments and statistical analysis. Naturally, measures of total work for
different numbers of cores are not comparable except when the benchmark
computation can be completed, so the same limitations apply here as above with
respect to test instances. One possible way to address this is to measure the
rate at which work can be preformed, as in the use of node throughput
described above.

\subsubsection{Measuring Changes in Progress \label{sect:changesProgress}}

Earlier in Sections~\ref{sect:scalability}, we proposed that reliable measures
of progress could ultimately lead to more robust techniques for measuring the
(weak) \Scal of parallel algorithms. Measures of progress are able to address
some of the difficulties inherent in measuring strong scalability because they
are meaningful at intermediate stages of computation. This eases the difficulty
of constructing a test set and allows assessment of algorithms that cannot
complete a given computation within a time limit.
To be suitable in the context of scalability, a measure of progress only needs
to be effective at measuring the fraction of a computation that has been
completed for computations that have already been previously attempted with
other resource configurations or other algorithms. Using information gathered
over repeated runs, we conjecture that it may be possible to compare progress
made by different algorithms or resource configurations. Thus, despite the
fact that none of the measures of progress developed thus far have proven
effective in other contexts, there is some reason to maintain optimism that
they will be effective in this context.

The optimality gap, while widely used as a measure of progress, seems to be a
difficult for for assessing \Scal, primarily due to the unpredictability of its
evolution. Subtle changes in timing could lead to an entirely different
assessment of progress. Gap integrals, on the other hand, may hold more promise,
as they are inherently more stable in their evolution. We explained earlier how
the gap integral could be used to compare the progress of two different
algorithms or two different resource configurations for solving the same
instance and this is exactly the kind of analysis that is needed to assess
scalability. Tree size estimates could similarly be used if any proved to be
accurate predictors, but currently this appears not have a very high potential.

\section{Summarizing and Visualizing \label{sect:aggregationAndVisualisation}}

The effective reporting of results is as important as the effective
performance of the experiments themselves, yet the importance of this step is
often overlooked. As we have already mentioned in
Section~\ref{sect:performanceAndScalability}, empirical experiments can be
thought of as methods of approximating the theoretical probability
distribution of all instances in a given class with respect to a given measure
of effectiveness. The goal in the reporting of results is to provide the
reader the data necessary to draw conclusions in a form that is easy to
consume. Typically, one wishes to determine which implementation will be most
advantageous for a particular use case. This comparison of implementations is
done essentially by comparing the empirical distributions produced by the
experiments. The two main tools for comparison are summarization (via the
computation of statistics associated with the distributions) and visualization
of the empirical CDFs themselves. Most techniques for doing this were
initially designed for the assessment of \Perf, but are also suitable for
assessing \Scal in general. Below, we discuss summarization and visualization
in the contexts of both efficiency and scalability analysis.

It is important to mention here once again that while the experiments are
intended to approximate the distribution of some measure of \Perf across an
entire class of instance, it is obvious that the quality of the approximation
is highly dependent on the instance set, as discussed in Section
\ref{sect:instanceSelection}. Ideally, conclusions regarding differences
between algorithms should be drawn using statistical methods and care must be
taken when choosing the instance set, which should be representative of all
instances from a class. Despite best intentions, the selection of instances will
more than likely introduce some unintended bias and it would be inaccurate to
state unequivocally that any of the methods below constitute a way of drawing
general conclusions beyond the extremely narrow setting of the experiments
themselves. 

\subsection{\PERF \label{sect:performanceSummary}}

\subsubsection{Summarization}

Historically, empirical data arising from solution of a set of test instances
was simply summarized and reported using a single summary statistic, usually
the arithmetic mean. The popularity of this method of summarization derives
from the fact that this statistic is simple to compute and the resulting
values appear meaningful (and sometimes are). However, it has been recognized
over time that a significant limitation of the arithmetic mean is that it can
be dominated by the results of only a few instances. As such, differences may
appear more significant than they actually are.

In an attempt to address the limitations of the arithmetic mean, the geometric
mean was proposed as an alternative for summarizing experimental results.
Unfortunately, the value of the geometric mean can also be dominated by a few
instances, particularly those with comparatively \emph{small} values.

The current standard for summarizing empirical results, first proposed
by~\mycitet{Achterberg}{achterberg07}, is the so-called shifted geometric
mean, computed by adding a positive scalar amount to each data point before
computing the geometric mean and then subtracting the same amount from the
result. By ``shifting'' the value, many of the issues associated with the
geometric mean can be alleviated, in particular where the reported value can
be dominated by few instances with very small values. The shifted geometric
mean is defined as follows.
\begin{definition}
  Given a set of values $N := \{x_{1}, x_{2}, \ldots, x_{n}\}$ and a shift value
  $s$, the \emph{shifted geometric mean} is given by
  \begin{equation*}
    SG(N) = \left(\prod^{n}_{k=1}(x_{k} + s)\right)^{\frac{1}{n}} - s.
  \end{equation*}
  \label{def:shiftedGeometricMean}
\end{definition}
The main motivation for using the shifted geometric mean is that test sets
typically contain instances of varying difficulty. For example, the wall clock
time required by \scip 3.2.1 (with \soplex 2.2.1) for solution of
the \miplib{}2010 instances ranges from 0.02 to 4414.33 seconds (excepting
those that take more than 7200 seconds). The shifted geometric mean is seen as
providing a more representative summary of results from such a diverse set of
instances, since it protects against bias from results that are both too large
and too small. 

The use of a single summary statistic has significant limitations for
reporting meaningful assessment results. If there are instances that fail to
solve within the time limit, then the use of summary statistics is seriously
hampered. When using means, geometric or arithmetic, this issue is commonly
addressed by computing the mean using only results from instances that are
solved within a given time limit and reporting the number that are unsolved.
However, this can clearly create a bias towards implementations that are highly
successful on some instances and fail completely on others. Alternatively,
this issue can be addressed by using a measure of \Perf that is better suited
for handling instances that fail to solve within the time limit (such as the
gap integral) or by using a different collection of test instances.

\subsubsection{Visualization}

Beyond single summary statistics, the use of graphical comparison methods have
gained significant traction in recent years and for good reason. Graphical
methods provide a more nuanced view of the behavior of implementation then
summary statistics, which greatly improves the ability to assess and compare
implementations. The visualization methods described below involve displaying
the empirical CDF of either the raw measures of \Perf
employed in the experiments or ratios of these measures (to provide a
scale-invariant comparison).

\paragraph{Performance Profiles.}

The performance profile, popularized by~\mycitet{Dolan and Mor{\'e}}{dolan02},
provides a visualization of the empirical CDF of the ratios of a given measure
of efficiency to the so-called ``virtual best'' for each algorithm and each
instance in a benchmark test set. Using a ratio to the virtual best rather
than the measure itself creates a scale invariant method of comparison and
therefore may avoid the potential bias introduced by the large differences in
resource requirements between instances that commonly arise in practice.

More precisely, the performance profile is based on the performance ratio
given by
\begin{equation*}
  r_{p, s} = \frac{t_{p,s}}{\min\{t_{p,s}\,|\,s \in S\}},
\end{equation*}
where $p$ is the instance, $s$ is the implementation and $t_{p,s}$ is
the measure of \Perf for $s$ in solving $p$. The profile for
implementation $s$ is then given by
\begin{equation*}
  \rho_{s}(\tau) = \frac{1}{|P|}\bigm|\{p \in P\,|\,r_{p,s} \leq \tau\}\bigm|,
\end{equation*}
where $\tau$ is a given performance ratio. The interpretation of
$\rho_{s}(\tau)$ is the fraction of problem instances that are within a ratio
of $\tau$ when solved by $s$. The function $\rho_{s}(\tau)$ is exactly the
CDF of the performance ratio.

\begin{figure}[htbp]
  \begin{center}
    \includegraphics[width=0.31\textwidth]{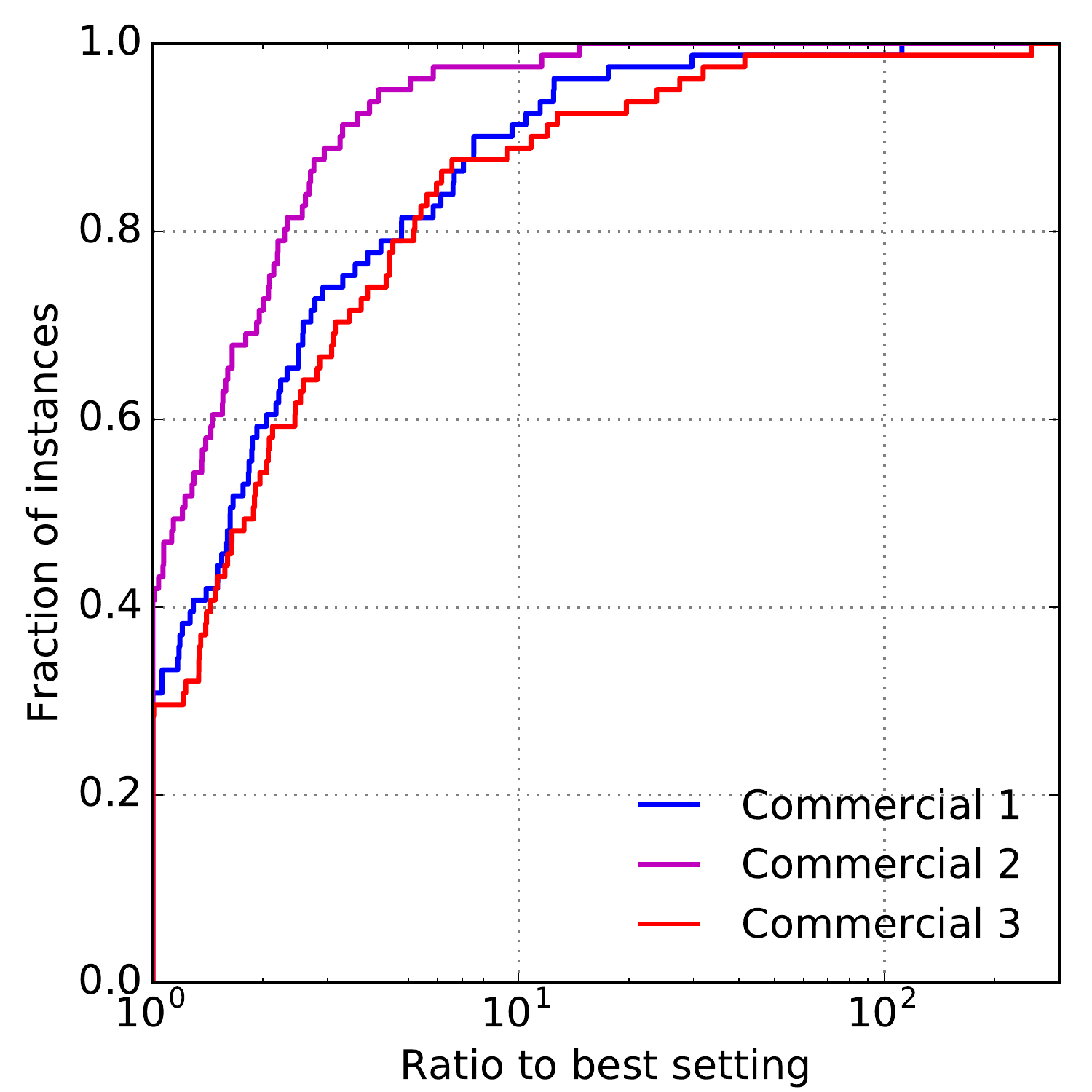}
    \caption{The performance profiles of the \wct for commercial solvers.}
    \label{fig:performanceWallclockCommercial}
  \end{center}
  \vspace{-5mm}
\end{figure}

An example of the use of performance profiles is shown in Figure
\ref{fig:performanceWallclockCommercial}, where the efficiency of three
commercial \mip solvers are compared on the \miplib{}2010 benchmark test set.
While performance profiles are often a very good tool for comparing
implementations, there are also limitations to their use, some of which are
highlighted by \mycitet{Gould and Scott}{gould16}.
Using the example in
Figure \ref{fig:performanceWallclockCommercial}, there seems to be a clear
dominance of \comb over \coma and \comc with respect to
\wct. Upon closer examination, however, it can be seen that the use of ratios
can still distort the results if the measures for some instances are close to
zero. For example, the difference between a solution time of .05 seconds and
0.2 seconds would probably be considered negligible to most, but the ratio of
the two running times is 4, which would be viewed as a very large ratio if
evaluated without knowledge of the underlying running times. For many of the
instances from the \miplib{}2010 test set, run times are very small for the
commercial solvers. As such, one would expect some large ratios to arise,
while practical performance of the solvers is actually very similar \Perf.
This issue could be remedied by using a different test set or by using a shift
value in the ratio calculation.

\begin{figure}[htbp]
  \begin{center}
    \subfigure[Excluding timeouts]{
      \includegraphics[width=0.31\textwidth]{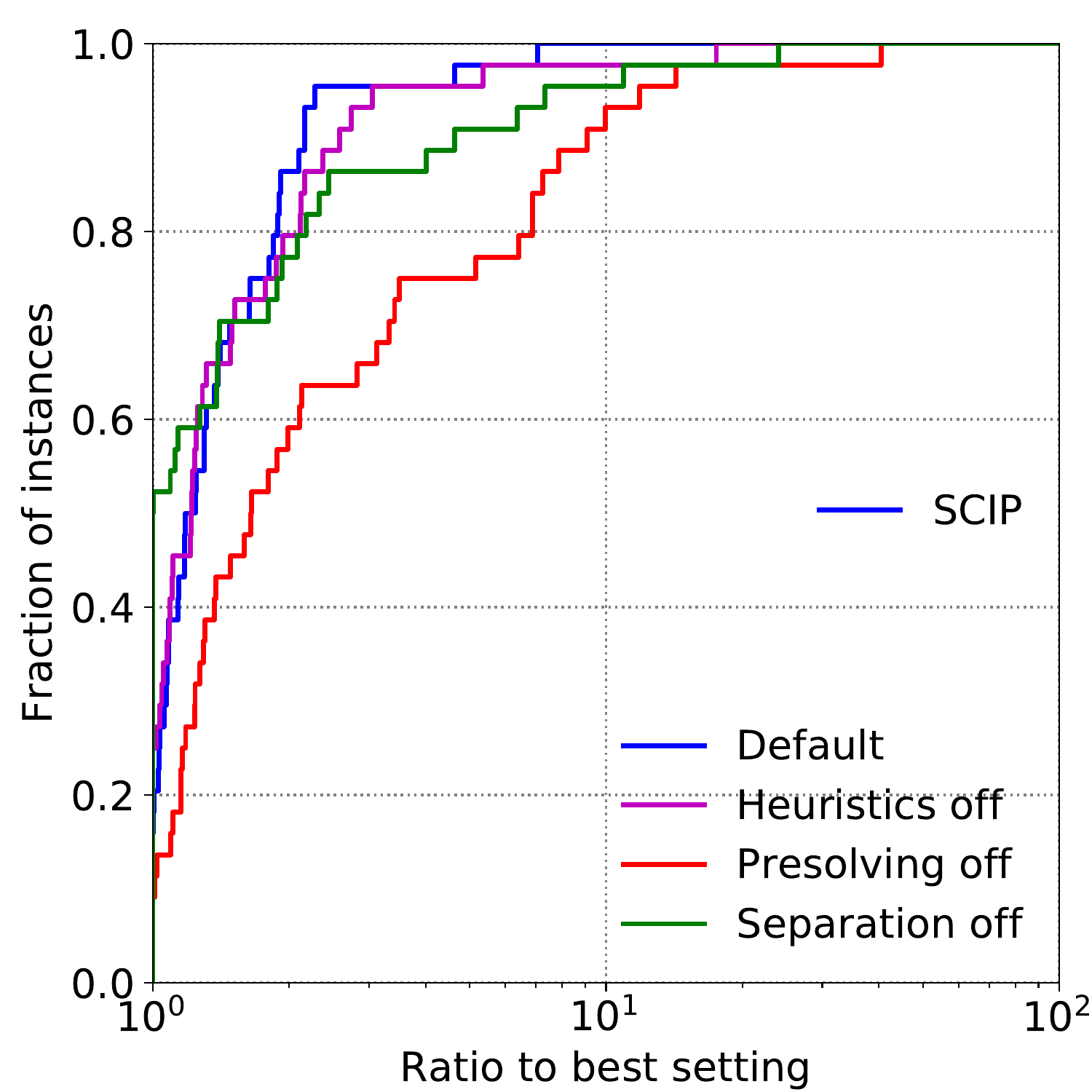}
      \label{fig:performanceProfileScipNotimeout}
    }
     \hspace{0.7in}
    \subfigure[Including timeouts]{
      \includegraphics[width=0.31\textwidth]{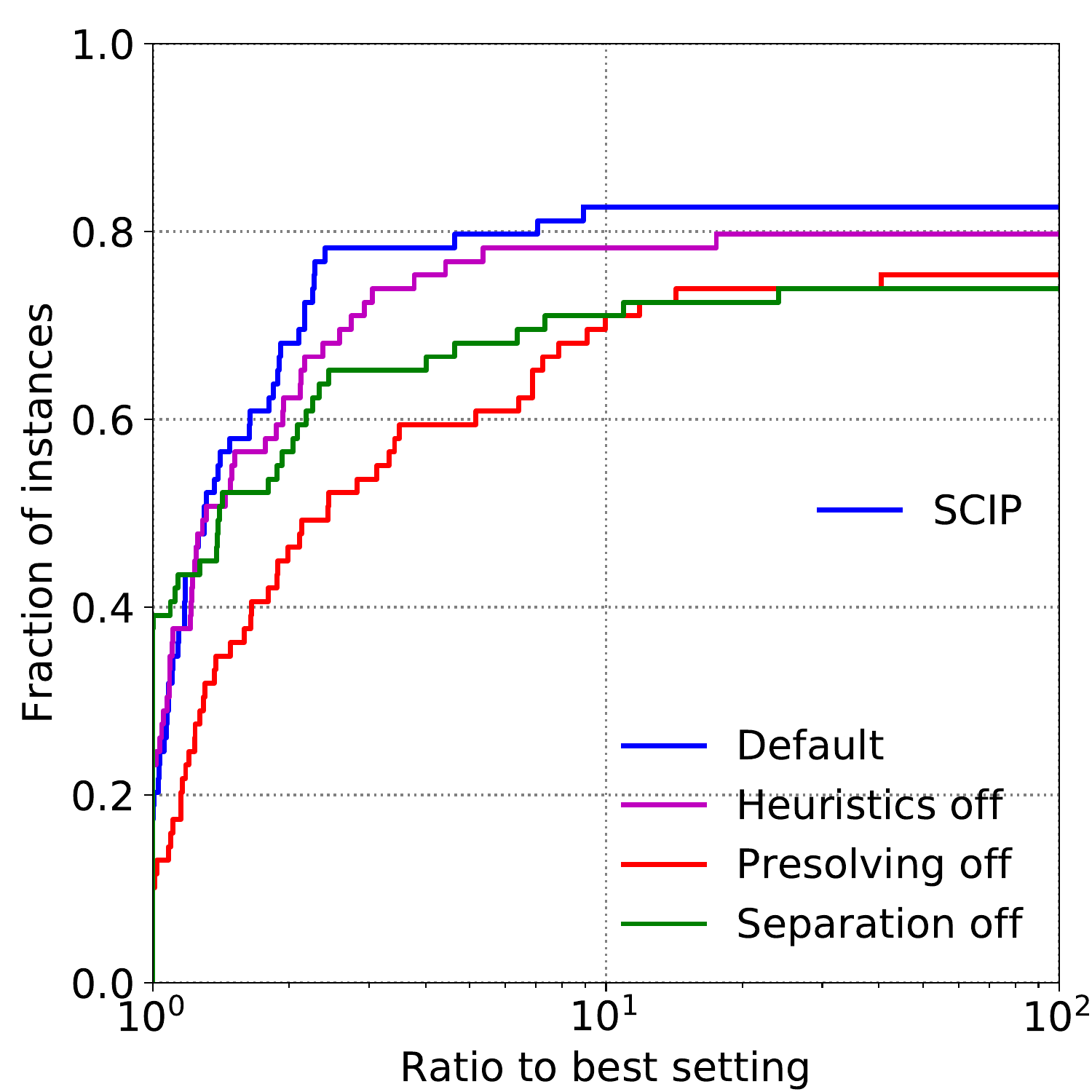}
      \label{fig:performanceProfileScipTimeout}
    }
    \caption{Performance profiles of the \wct when executing \scip with different emphasis settings}
  \label{fig:performanceProfileScip}
  \end{center}
  \vspace{-5mm}
\end{figure}

Another limitation arises when the test set contains instances that fail to
solve within the time limit. In this case, the classical measure of \wct is no
longer applicable.
While performance profiles can visually indicate the fraction of instances for
which a given implementation did not complete the solution process within the
time limit, it cannot give any indication of the progress made on those
instances. Including the instances that did not complete in the profile makes
the profile more difficult to read while adding very little additional
information. Two different performance profiles are presented in
Figure \ref{fig:performanceProfileScip}, both evaluating the \Perf of \scip
when solving instances from the \miplib{}2010 benchmark test set using
different meta-settings.
The parameter settings used for the computational
experiments, in addition to the default settings, disable the heuristics,
separation or presolving, labeled as \emph{Heur off}, \emph{Sepa off} and
\emph{Presol off} respectively. In Figure
\ref{fig:performanceProfileScipNotimeout}, only the instances that did not time
out or abort for all meta-settings are included.  Alternatively, Figure
\ref{fig:performanceProfileScipTimeout} displays the performance profile using
all instances, so instances that time out are included. The two profiles
display the same information, but in the second one, the intercept on the
right side is lower to indicate the percentage of instances that timed out.

\paragraph{Cumulative Profiles.}

As we have seen, one of the main appeals of performance profiles is the
unitless comparison that results from the use of ratios of a given measure
across a set of implementations. While the unitless comparison is valuable, it
is not suitable in all contexts.
An alternative to the performance profile is the cumulative profile, which is
a plot of the fraction of instances in a test set solved within a given time
limit and thus directly visualizes the empirical CDF. More precisely, let
$t_{p, s}$ denote  the measure of \Perf associated with instance $p$ and
implementation $s$. Then, the cumulative profile for implementation $s$ is a
graphical representation of the function \begin{equation*} f_{s}(t) =
\frac{1}{|P|}|\{p \in P\,|\,t_{p,s} \leq t\}|, \end{equation*} which is the
empirical fraction of instances solved within $t$ seconds.

\begin{figure}[bp]
  \begin{center}
    \includegraphics[width=0.31\textwidth]{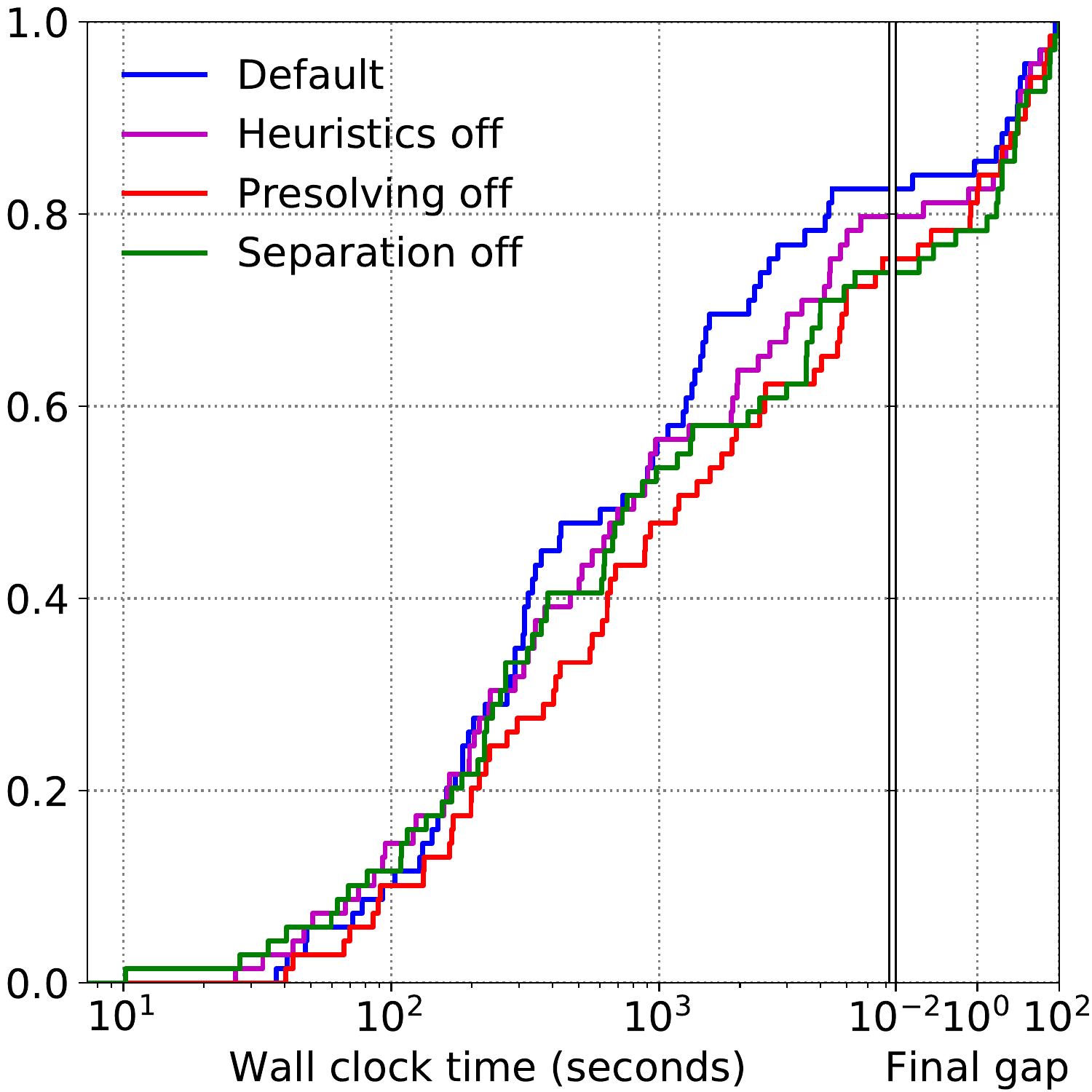}
    \caption{A cumulative profile combining the \wct and final gap when running \scip with
  different emphasis settings} \label{fig:profileScip}
  \end{center}
  \vspace{-5mm}
\end{figure}

The cumulative profile provides a different perspective on the assessment
of \Perf and \Scal than the performance profile. In particular, the cumulative
profile is a representation of the \Perf of each algorithm in an absolute rather
than relative sense. Although the cumulative profile apparently suffers from the
same difficulty in dealing with instances that fail to solve within the given
time limit, this issue can be mitigated by extending the cumulative profile by
utilizing a measure of \Perf or progress that applies to instances that do not
complete within the time limit. For instances, it is possible to combine the
cumulative profile for the solved instances with the cumulative profile of
the final gap for the unsolved instances.  Figure \ref{fig:profileScip} presents
such an example using the same data that produced Figure
\ref{fig:performanceProfileScip}.   A different example of a cumulative profile
is presented by~\mycitet{Dinh et al.}{dinh18}.

The combination of the measure of \Perf and the final gap attempts to provide an overall
assessment of the test set when instances time out. However, some of the
limitations of both measures are highlighted in Figure \ref{fig:profileScip}.
First, only a subset of instances are assessed for each setting in the left hand
plot---this is potentially different subset for each setting. Thus, the instances
used to produce the \wct and final gap plots could be completely different for
each of the evaluated settings. A further limitation of this particular figure
is related to the final gap. Specifically, the final gap is bounded---namely to
100\%---as given by equation \eqref{eqn:primaldualgap}.  This is not necessarily
a limitation on all gap computations, but the fact that there are many ways to
compute the gap is an issue that needs to be considered when using this measure.

\subsection{\SCAL \label{sect:scalabilitySummary}}

\SCAL analysis assesses the \emph{change} in efficiency that occurs as the
bundle of available resources are changed. Thus, we must consider not just a
single measure, but a trade-off, which gives the data has an additional
dimension. Although this apparently demands new methods of summarization and
visualization, we can adapt the methods discussed in the previous section to
some extent by simply accounting for the extra dimension in the data. Since
analyzing \Scal involves comparing the \Perf of a single implementation as
resource levels are varied, we can in principle apply the methodology for
comparing the \Perf of different implementations from the previous section to
that of the same implementation with differing bundles of resources.  However,
we face some challenges in doing this that we'll describe below and whose
resolution is the subject of on-going work. Some work on visualization of
scalability data has already been done by~\mycitet{Koch et al.}{koch12}, who
present methods for visualizing the parallel efficiency, changes in node
throughput, and changes in number of nodes processed for individual instances,
However, these were for the analysis of single instances and do not constitute
methods of summarization across a test set.
We describe here only more traditional methods based on strong scalability
here.

\subsubsection{Summarization}

The experimental data produced when analyzing \Scal is similar to that
produced when analyzing \Perf, but we have a separate set of data for each
resource level tested. For example, we may measure the efficiency across a
test set when given 1, 4, 16, and 64 cores. The data collected from the
experiments with 4, 16 or 64 cores are of precisely the same nature as that
collected for 1 core.

Summarization of \Scal data involves both summarization across the test set
(within a single experiment) and summarization across multiple experiments
involving a single solver. Although researchers have tended to summarize the
data within an experiment first and then compute \Scal statistics based on
those summary statistics, richer analysis may be possible when \Scal
statistics are computed for each instance in each experiment and then these
statistics are in turn used to assess and compare the scalability of different
implementations. By considering the speed-up or efficiency on a per-instance
basis, we may be able to more easily compare the scalability of multiple
parallel implementations by mitigating the effect of the inherent differences
in the potential for scalability of individual instances. For example, we
could visualize the parallel efficiency of the solution process for individual
instances using a performance profile to compare implementations.

\subsubsection{Visualization}

\paragraph{Speed-up Plots.}

\begin{figure}[bp]
  \begin{center}
    \subfigure[\Wct{}, with 5 hr limit]{
      \includegraphics[width=0.31\textwidth]{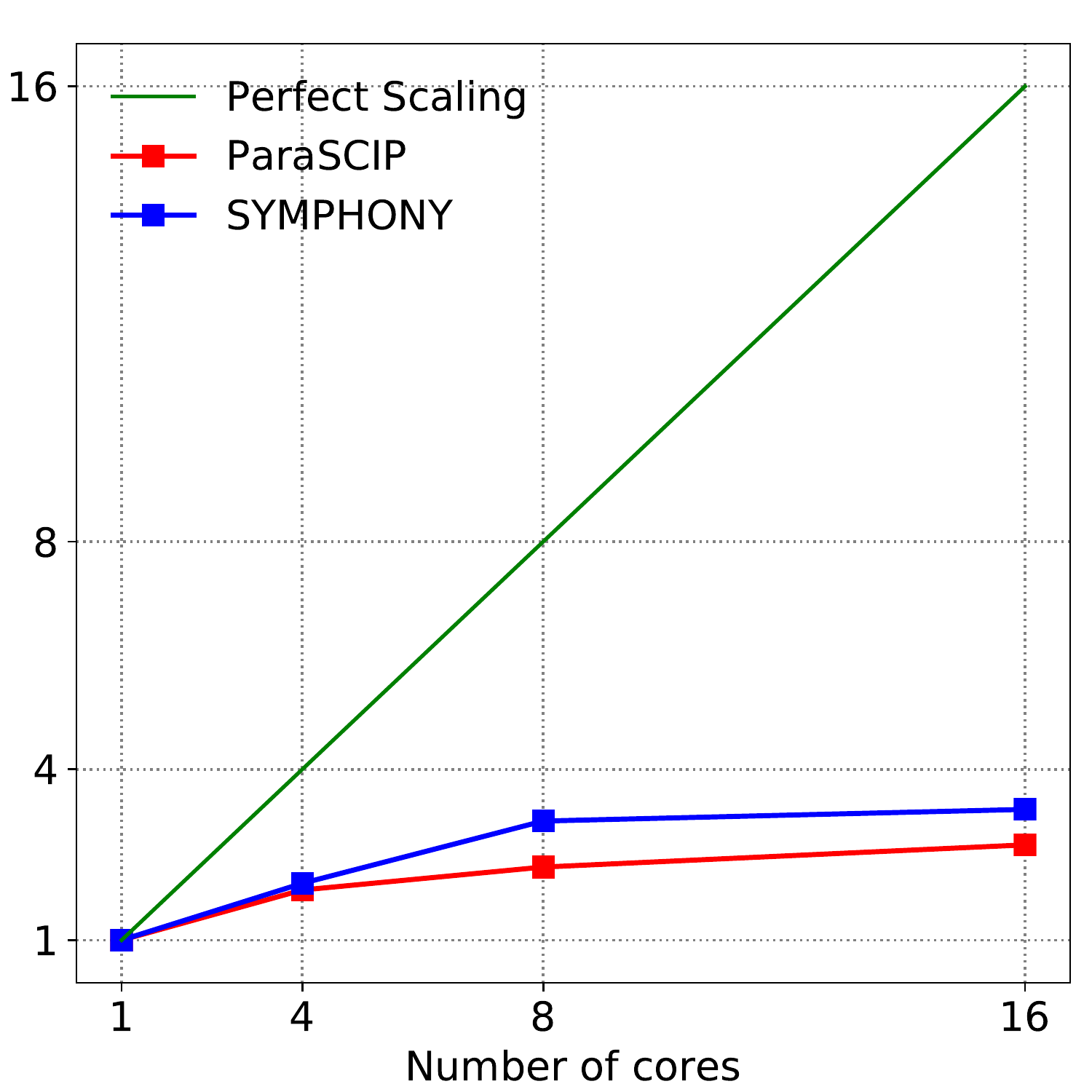}
      \label{fig:parascipSymphonySpeedupWct} }
      \hspace{0.7in}
    \subfigure[\Pdi{}, all instances]{
      \includegraphics[width=0.31\textwidth]{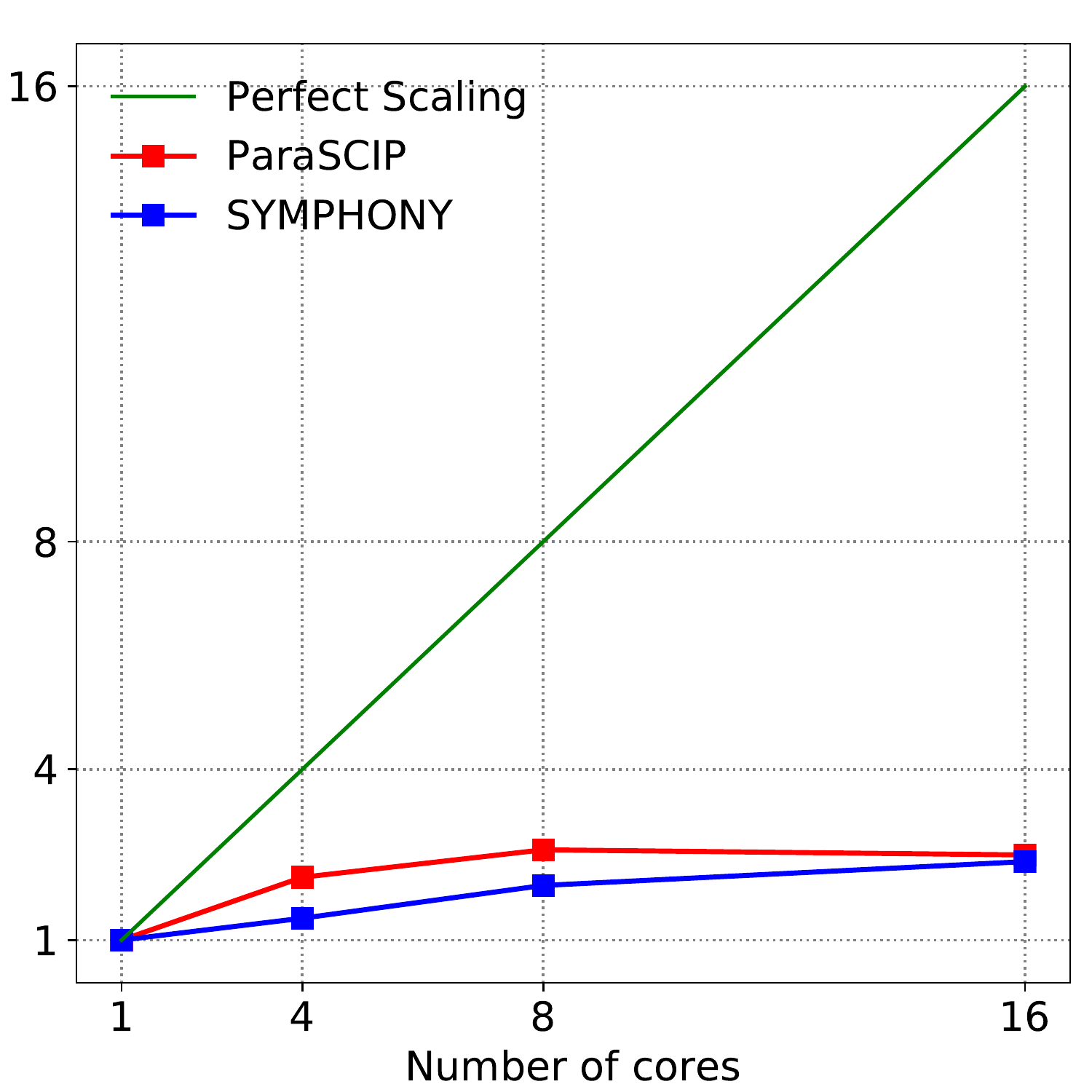}
      \label{fig:parascipSymphonySpeedupPdi} }
    \caption{The speed-up of \parascip and \symphony based of the wall clock
    time and \pdi, using the shifted geometric mean on instances from \miplib{}2010 that did not abort.}
  \label{fig:parascipSymphonySpeedup}
  \end{center}
  \vspace{-5mm} \end{figure}

The most commonly used method of presenting scalability data applies to
experiments in which the \wct is fixed and the number of cores is varied.
First, summary statistics are computed for each benchmark computation and
speed-up is then computed for each number of cores and each implementation.
Finally, a plot similar to that in Figure \ref{fig:parascipSymphonySpeedup} is
displayed graphically comparing implementations to each other and to the
reference line for the ideal case of linear speed-up).  Figure
\ref{fig:parascipSymphonySpeedup} compares \parascip and \symphony on instances
from the \miplib{}2010 benchmark test set for which neither solver aborted and
(for Figure \ref{fig:parascipSymphonySpeedupWct}) that were solved within 5
hours by both solvers across all resource configurations.

Using Figure \ref{fig:parascipSymphonySpeedup} as an
example, we can clearly see that as the number of cores increases, the marginal
speed-up that is achieved for both \parascip and \symphony decreases. In
comparison to perfect scaling, which is given by the green line, this figure
suggests that there is very little return from increasing the number of cores
beyond 16.

Figure \ref{fig:parascipSymphonySpeedup} encapsulates many of the challenges
that have been discussed in the preceding sections. First, the speed-up of a
parallel implementation is computed from the summary of an underlying measure.
In Figure \ref{fig:parascipSymphonySpeedupWct}, the underlying measure is the
\wct which has been stressed as a problematic measure for \Perf and \Scal
assessment. In this example, because \wct was used, it was only possible to
compute the speed-up on the instances that could be solved within 5 hours.
Thus, many instances have been discarded from the analysis, which biases the
results. In particular, the instances discarded are those that are the most
difficult, which also tend to be those for which it is easier to achieve \Scal
and that may benefit the most from parallelization. When comparing \Scal of
different implementations, the issue is even worse, since we must discard
instances not solvable by any of the implementations. One possible fix is to use
a measure such as the \pdi that does not require all instances to be solved
within the specified time limit. To illustrate, Figure
\ref{fig:parascipSymphonySpeedupPdi} shows the ``speed-up'' calculated from the
\pdi rather than \wct. Calculating a speed-up based on \pdi may seem rather odd,
but the obvious advantage of using a gap integral as a measure of \Perf is that
the full test set can be utilized and the results are not biased by the
existence of instances that cannot be solved by one of the implementations being
compared.

A second issue is that the \Scal of a parallel implementation is evaluated by
using a specific test set. In the case of Figure
\ref{fig:parascipSymphonySpeedup}, the instances used were from the
\miplib{}2010 benchmark test set. The result presented in Figure
\ref{fig:parascipSymphonySpeedup} suggest that both \parascip and \symphony fail
to scale very well with an increase in the number of cores.  However, this poor \Scal
may not be due to the parallel implementation but associated with the test set
used to do the evaluation. As explained in Section
\ref{sect:instanceSelection}, if the sequential algorithm is able to solve the
instances in a small number of \bnb nodes, then it is almost guaranteed that a
parallel \bnb implementation will be unable to achieve a \Scal close to perfect.
Thus, the evaluation of \Scal from the speed-up plot requires a considered
choice about the test set used for the computational experiments.

Finally, the speed-up plot only provides a very high level overview of
the \Scal of parallel implementations. The only analysis that can be performed
from such a visualization is to assess how different the \Scal of an
implementation is from perfect scaling across the entire test set. It is not
possible to gain a better understanding of the implementation and identify how
to improve the \Scal or to understand how scalability varies from instance to
instance. This method of visualization should only be used to present a
high-level summary, but should generally be backed up by other
summary and visualization techniques.

\paragraph{Performance Profiles.}

A more nuanced way of visualizing scalability data may be obtained by using
performance profiles. The use of a performance profile for assessing \Scal may
at first seem a little strange, but there are actually multiple ways in which
they can be applied and the resulting analysis is richer than what can be
derived from a simple speed-up plot. The use here differs subtly from the
typical use of this visualization technique in assessing \Perf.

\begin{figure}[bp]
  \begin{center}
    \subfigure[\parascip - excluding timeouts]{
      \includegraphics[width=0.31\textwidth]{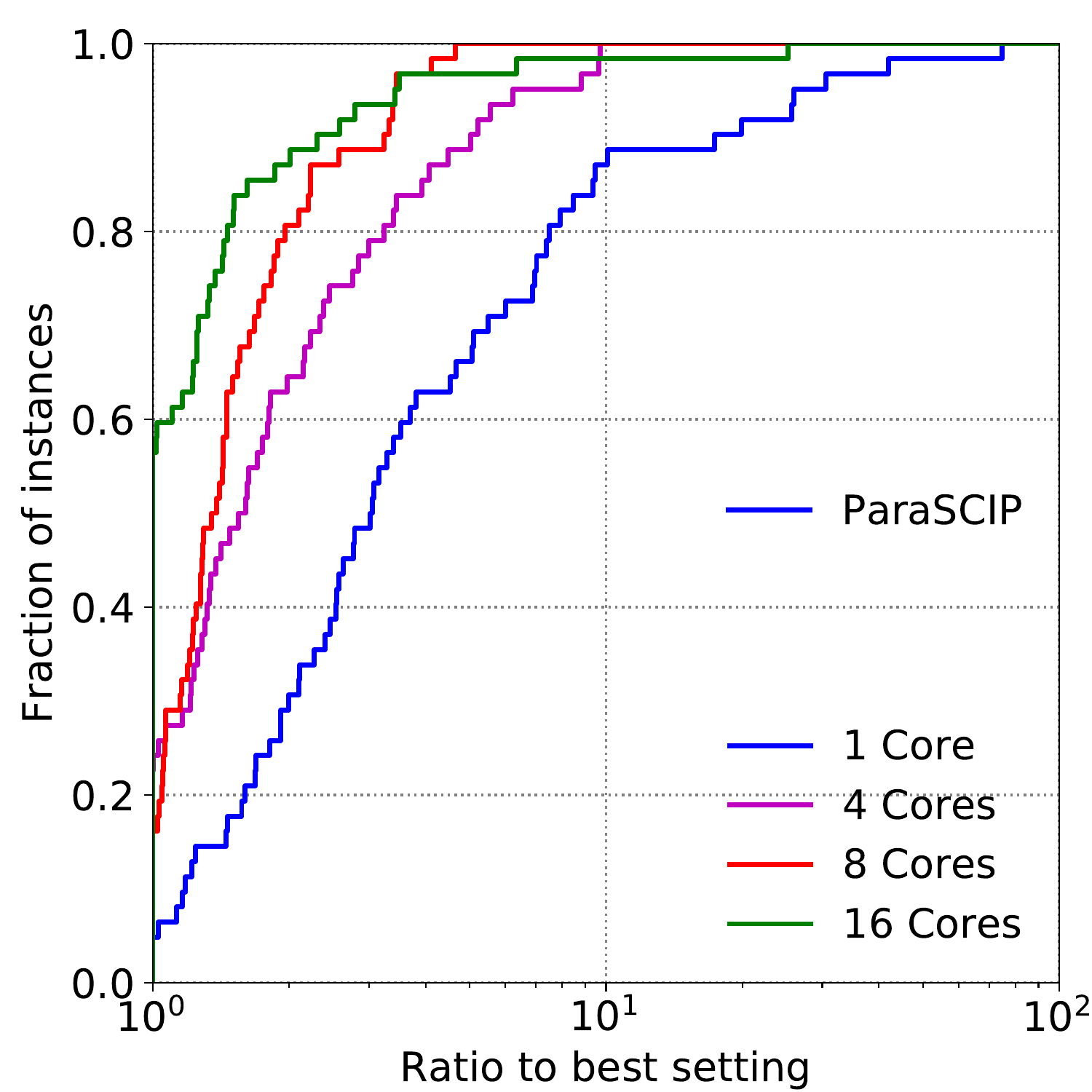}
      \label{fig:performanceWallclockParascipNotimeout}
      }
       \hspace{0.7in}
    \subfigure[\symphony - excluding timeouts]{
      \includegraphics[width=0.31\textwidth]{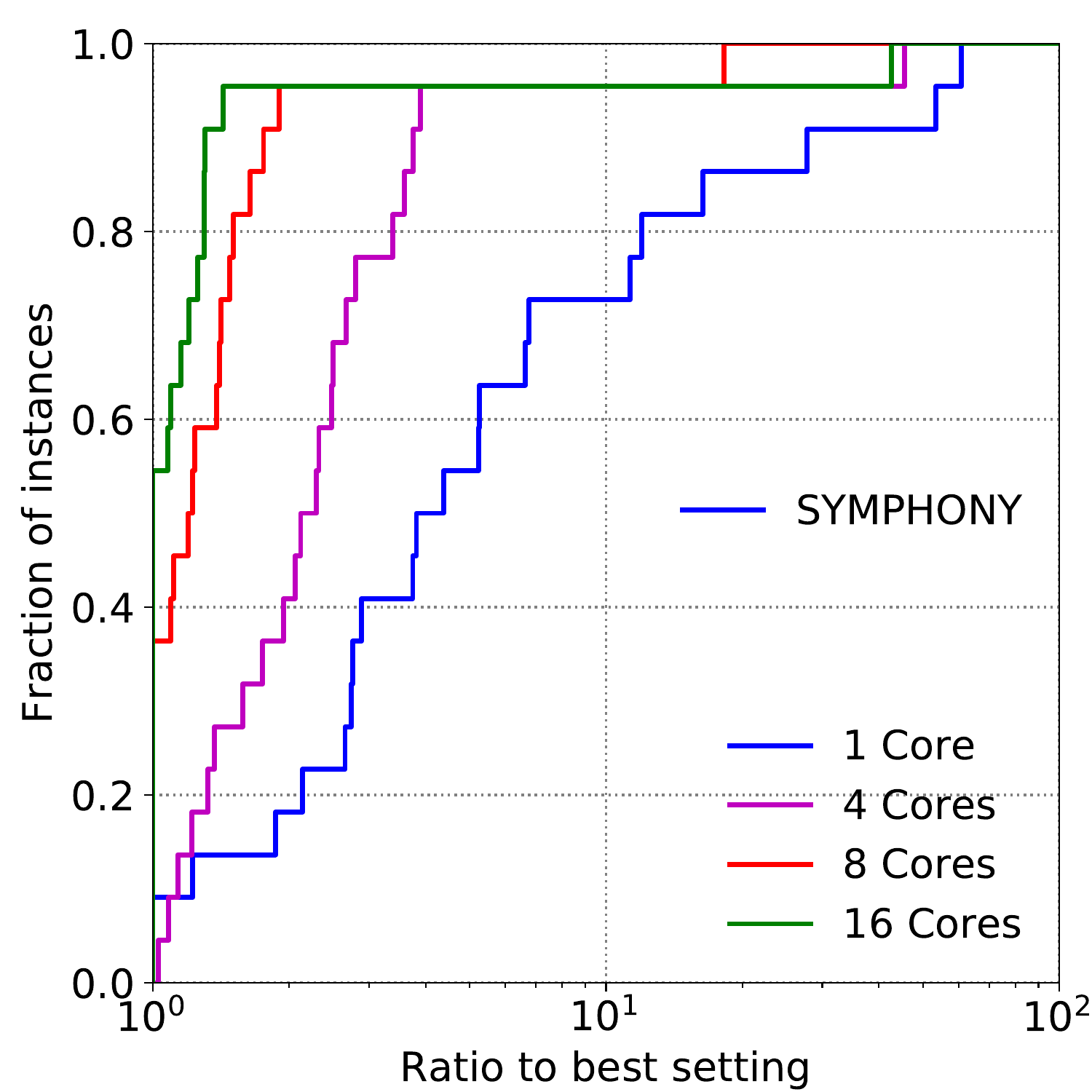}
      \label{fig:performanceWallclockSymphonyNotimeout}
      }\\
    \subfigure[\parascip - including timeouts]{
      \includegraphics[width=0.31\textwidth]{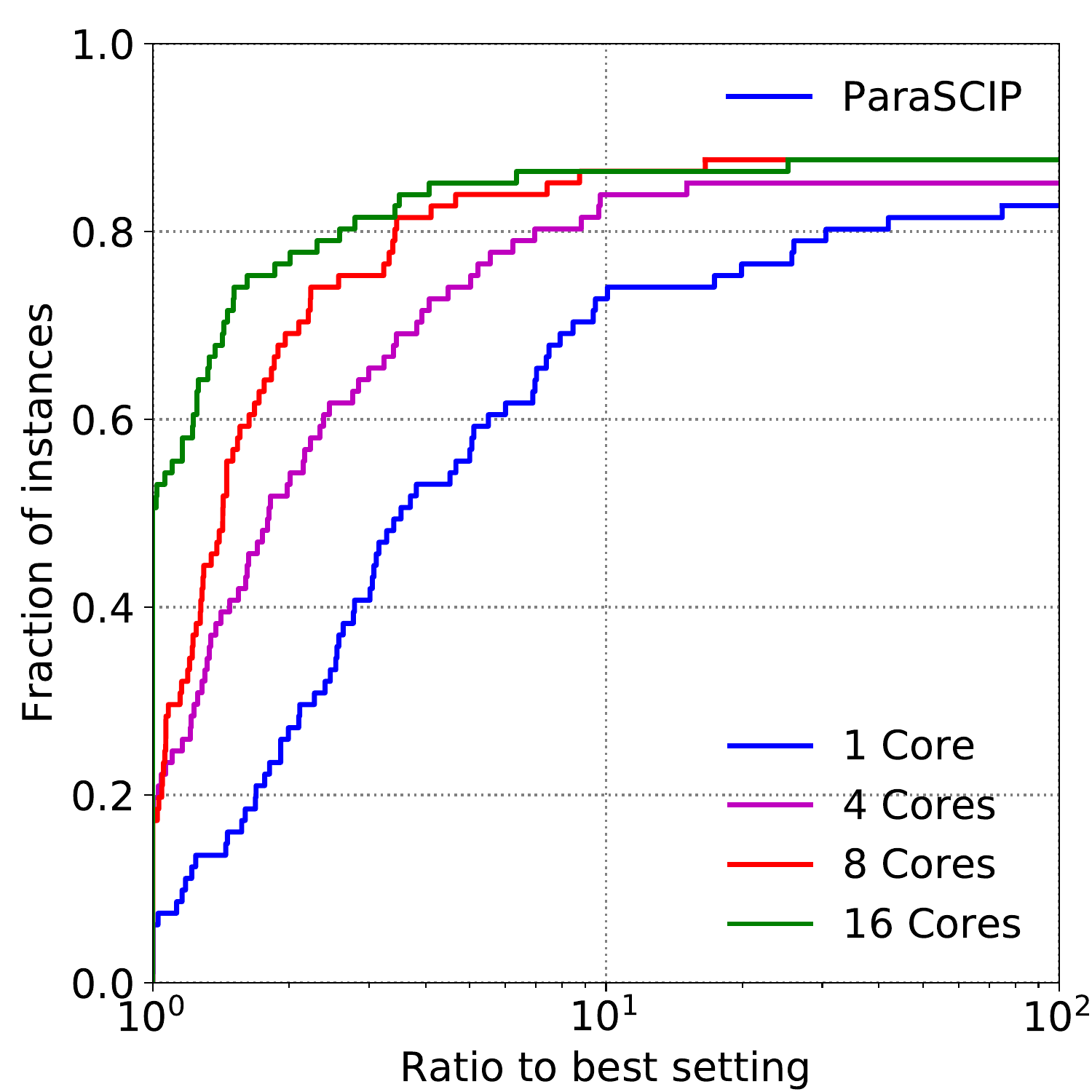}
      \label{fig:performanceWallclockParascip}
      }
       \hspace{0.7in}
    \subfigure[\symphony - including timeouts]{
      \includegraphics[width=0.31\textwidth]{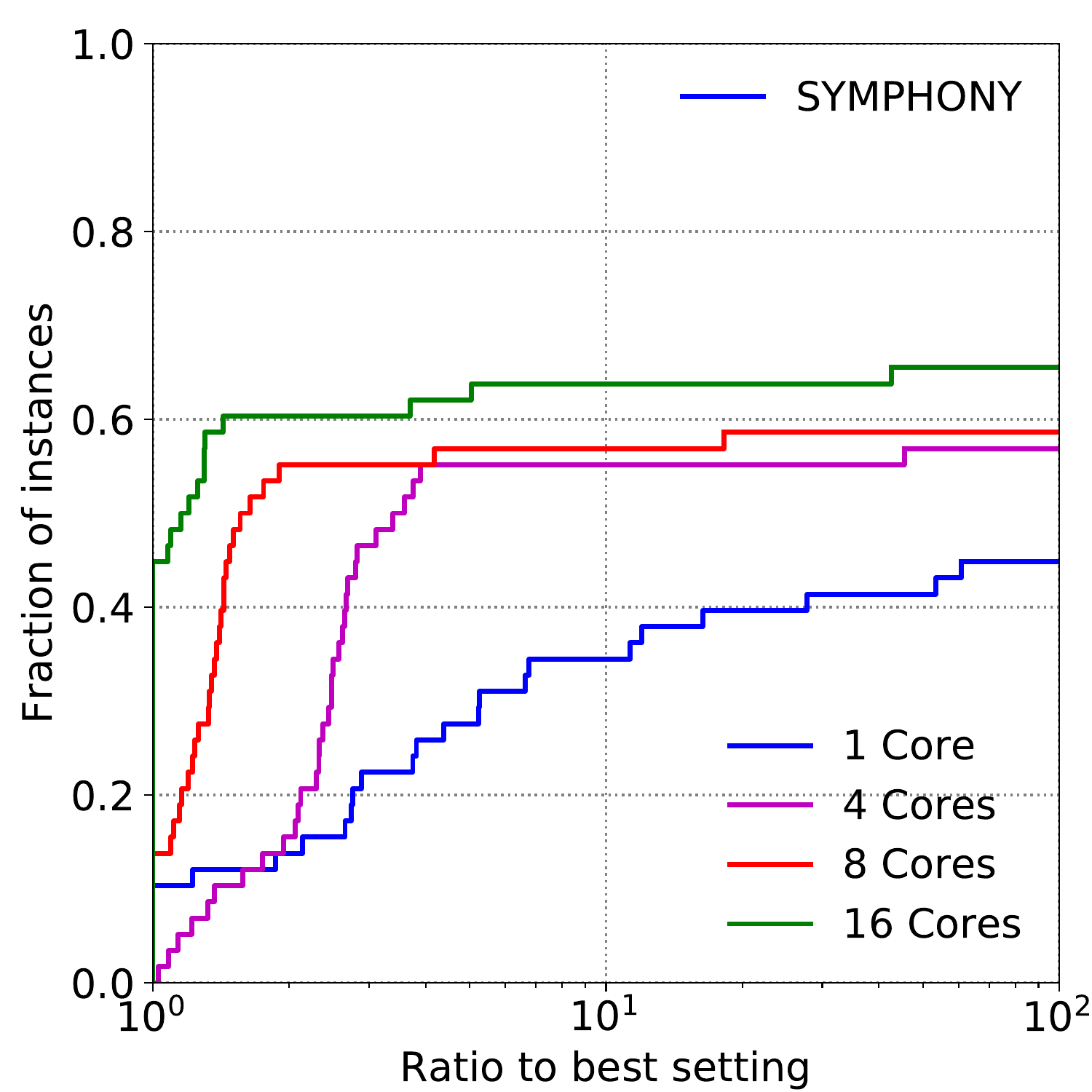}
      \label{fig:performanceWallclockSymphony}
      }
    \caption{The performance profile of the \wct for \parascip and \symphony using different numbers of
    cores.}
    \label{fig:performanceWallclockParascipSymphony}
  \end{center}
  \vspace{-5mm}
\end{figure}

While performance profiles are generally used to compare different
implementations, here we suggest using them to assess the \Scal of a single
underlying implementation. In this context, the comparison is with respect to
different resource configurations, typically changes in the number of cores. An
example of using performance profiles to compare the parallel implementations of
\parascip and \symphony with different numbers of cores is presented in Figure
\ref{fig:performanceWallclockParascipSymphony}.

One can clearly see the expected improvement in the measure of \Perf as the
number of cores are increased, but the results with different numbers of cores
are not directly comparable, since more total core hours are allocated to the
computations with more cores. This may make it less obvious that there are
actually efficiency \emph{losses} as the number of cores is increased if one
considers the progress made per core hour allocated to the computation. 

A more insidious issue with using performance profiles for \Scal assessment,
however, is related to computation of ratios with respect to the ``best''
implementation for each instance. While it is expected that the implementation
with the greatest number of cores will be the ``best'' implementation, this
may not be the case for all instances of the test set. As demonstrated in
Figure
\ref{fig:performanceWallclockParascipSymphony}, the 16-core implementation
appears to be the ``best'' implementation for approximately 50\% of the
instances for both \parascip and \symphony. Even the sequential implementation
is the ``best'' implementation for about 10\% of the instances for both
implementations. This variation in the ``best'' implementation suggests that for
individual instances the \Scal of the parallel implementations is not perfect.
While we don't expect perfect \Scal for individual instances, the average
scaling across the complete test set is difficult to evaluate from this
visualization technique alone.  Unfortunately, assessing whether an
implementation scales well is obscured by the variability in the scaling for
individual instances.

In regards to assessing \Scal, performance profiles seem to introduce more
confusion than insight and there is clearly a need for alternative methods.
We propose that some of the issues with performance can actually be fixed, but
we leave discussion of that to a future work. 

\section{Concluding Remarks \label{sect:concludingRemarks}}

Assessing the effectiveness of \bnb implementations is an immensely difficult
task.  Since modern \bnb implementations comprise of numerous algorithmic
components, it is difficult to prescribe general methods for the effective
assessment of \Perf and \Scal.  Within this paper, we propose an assessment
framework that is based upon resource consumption. The assessment of \Perf
involves assessing the consumption of a single resource, typically time, while
keeping all other resources, such as memory and cores, fixed. While \Perf is
usually associated with sequential implementations, this concept of resource
consumption can equally be applied in the parallel context. \SCAL is a concept
related to parallel \bnb implementations that is evaluated as the trade-off
between two or more resources. The resources typically evaluated in a \Scal
analysis are time and cores.

Starting from this general framework of resource consumption, we highlight many
challenges and limitations of the current best practice for the assessment of
\bnb implementations. The sophistication of implementations, the
non-deterministic run time behavior---especially in the context of parallel \bnb
implementations---and performance variability all represent significant
challenges that must be considered when conducting empirical studies of
algorithms. Further, the classical measures of \Perf, work and progress exhibit
significant limitations. This is shown to be further accentuated when assessing
\Scal. 

This paper aims to initiate further discussion regarding the effective
assessment of \bnb implementations. We are building on a number of previous
works calling for improved standards and guidelines for empirical studies of
algorithms~\mycitep{barr93,beiranvand17,crowder78,hoffman53,hooker95,hooker94,ignizio71,jackson90,kendall16,Joh02,Gre90,McG87,McG12,MorSha01,Mor02,CofSal00}.
By highlighting the challenges of assessing \bnb implementations, we hope to
drive innovation in the development of effective assessment techniques. There
are many advancements that can be made in devising improved methods for the
selection of implementations and test instances. An example of developing
improved methods for test instance selection is the long running \miplib
project~\mycitep{achterberg06,bixby92,bixby98,koch11}. A large impediment to the
assessment of \bnb implementation is the prevalence of performance variability.
A better understanding of performance variability and improved strategies to
minimize its impact on algorithm evaluation will make great strides to a more
rigorous assessment of \Perf and \Scal.

\section*{Acknowledgments}

This work has been partly supported by the Research Campus MODAL
\emph{Mathematical Optimization and Data Analysis Laboratories} funded by the
Federal Ministry of Education and Research (BMBF Grant~05M14ZAM). SJM is
supported by the Engineering and Physical Sciences Research Council (EPSRC)
grant EP/P003060/1. 


\bibliography{scaling}
\bibliographystyle{abbrvnat}

\end{document}